# Arbitrary Total Angular Momentum Vectorial Holography Using Bi-Layer Metasurfaces

*Joonkyo Jung*[1], *Hyeonhee Kim*[1] and *Jonghwa Shin*[1,*]

[1]Department of Materials Science and Engineering, KAIST, Daejeon 34141, Republic of Korea

[*]E-mail: qubit@kaist.ac.kr

**Abstract:** Advanced holographic techniques are increasingly demanded for high-capacity and secure information processing. In this context, orbital angular momentum (OAM) stands out as a powerful resource for optical multiplexing, offering access to an unbounded set of orthogonal modes. To harness this potential, metasurfaces, with their considerable ability to control light, have emerged as key platforms for OAM-multiplexed holography. Nevertheless, conventional OAM holography suffers from limited polarization engineering capabilities due to the lack of chirality control in single-layer metasurfaces. Here, we introduce a bi-layer metasurface architecture that realizes total angular momentum (TAM) vectorial holography, where TAM represents the combination of spin angular momentum (SAM, equivalent to polarization) and OAM of light. In contrast to previous approaches, this scheme enables true polarization–OAM multiplexing, facilitating the independent generation of vectorial holographic images for each orthogonal TAM input state. This concept is validated numerically and experimentally, confirming the feasibility of TAM vectorial holography. The proposed scheme can be easily integrated with other recent holography generation approaches, such as vector beam multiplexing and bidirectional holography, thereby further expanding its multiplexing capability. This work establishes a versatile framework for advanced full-vectorial holography, showing how metasurfaces can unlock multiplexing strategies for emerging photonic systems.

**Keywords:** metasurfaces, total angular momentum, vectorial holography, polarization multiplexing, orbital angular momentum multiplexing, vector beam multiplexing, asymmetric transmission

## 1. Introduction

Optical holography offers a powerful means of recording and reconstructing optical fields while preserving their exact properties, including amplitude, phase, and polarization. This enables highly realistic three-dimensional displays as well as high-density optical data storage [1-3]. With rising demand, advanced holographic techniques—particularly optical multiplexing, the intentional encoding of independent information channels across multiple physical degrees of freedom of light—have emerged as key research directions for increasing data capacity, reducing device size, and inducing new functionalities [4-13]. The applications of optical multiplexing extend beyond holography, driving breakthroughs in high-speed optical communications [14, 15], quantum information processing [16, 17], and optical computing architectures [18, 19], establishing it as a vibrant research frontier across photonics.

Among the potential degrees of freedom that can be harnessed, orbital angular momentum (OAM), characterized by a helical phase front, has attracted considerable interest due to its unique property of supporting an unbounded set of orthogonal helical modes, thereby providing a promising pathway toward highly multiplexed holographic systems [20-22]. In this scheme, different sets of information are encoded according to the helical mode index, which specifies the number of $2\pi$ phase twists around the beam axis, allowing the reconstruction of independent optical fields for each OAM of the incident beam. Despite this intriguing idea, conventional bulk optics and spatial light modulators utilized for OAM holography impose limitations on device performance due to their inherent technological hurdles. Notably, their low spatial resolution restricts the achievable fidelity and channel density of OAM-multiplexed holograms [20]. Moreover, combining OAM with other optical degrees of freedom, such as the polarization states of incident and outgoing light, remains challenging.

To overcome these limitations, metasurfaces—planar arrays of nanostructures with subwavelength dimensions—have recently emerged as promising platforms for next-generation optical devices [23, 24]. By individually modulating the shape and dimensions of the constituent nanostructures, metasurfaces can tailor their optical responses, such as amplitude gains, phase delays, and polarization conversions, with subwavelength spatial resolution. In addition, they can be designed to behave differently depending on the configuration of the incident light [24-29], offering various forms of optical multiplexing at minimal additional cost. This versatility renders metasurfaces a highly adaptable platform for further advancing OAM holography.

In particular, the arbitrarily designable anisotropy of metasurfaces provides high degrees of freedom in polarization control and has given rise to two important subfields of OAM holography: polarization-multiplexed OAM holography [30-35] and OAM vectorial holography [36]. In polarization-multiplexed OAM holography, different holograms can be realized depending on the incident polarization and OAM, enabling the production of holographic images with designable intensities. In OAM vectorial holography, vectorial holographic images with engineered intensity and polarization distributions can be reconstructed for a given OAM input, as schematically represented in Figure 1. Nevertheless, current implementations based on single-layer metasurfaces face intrinsic limitations in polarization manipulation because they lack the controllability of the chirality of the system. The polarization-multiplexed OAM hologram can match the target holographic images only in its intensity pattern and not in its polarization distribution, whereas the OAM vectorial hologram can be designed only for a single input polarization at a given OAM. In other words, one must choose either the polarization-multiplexing capability or full-vector designability, but not both. Importantly, this intrinsic limitation cannot be overcome simply by changing the shape or dimensions of the nanostructures. Unifying these two subfields and realizing polarization-multiplexed, fully vectorial OAM holography mandates the introduction of new principles.

In this study, we propose a new approach to realize arbitrary total angular momentum (TAM) vectorial holography, in which each TAM state, defined by a specific polarization state (i.e., spin angular momentum, SAM) and an OAM mode of the incident light, is mapped to a distinct output vector field as represented in Figure 1. This is achieved by utilizing our recently proposed bi-layer metasurface architectures, which can overcome the symmetry constraints of conventional single-layer designs, thereby enabling complete control over anisotropy and chirality independently [37]. We numerically and experimentally demonstrate the proposed fully polarization–OAM-multiplexed TAM vectorial holography principle with examples. In addition, we propose two new holographic configurations enabled by integrating our proposed scheme into other state-of-the-art holographic multiplexing mechanisms: (1) vector beam-multiplexed vectorial holography and (2) bidirectional TAM vectorial holography. Our results establish a generalizable platform for TAM-multiplexed vectorial holography, paving the way toward TAM-enabled, highly multiplexed optical devices.

## 2. Results and Discussion

### 2.1. Arbitrary TAM vectorial holography

In contrast to conventional holography, where the illumination beam is typically a simple plane wave, OAM-"preserving" holography can utilize complex illuminations with non-trivial OAM; it can then reproduce the incident beam's OAM characteristics at each reconstructed image pixel, which is discretely sampled over the target image in a regular grid pattern. A related, but different configuration that embeds the conjugate phase of an input beam with the target OAM enables OAM-"selective" holography, where the metasurface cancels the helical phase of the target input beam and generates fundamental Gaussian-like beams with zero OAM that have local intensity maxima at the center of each image pixel. Since other input beams with different OAM are converted by the metasurface to beams with non-zero OAMs and a singular point (zero intensity) at the center of each pixel, simple spatial filtering with a pinhole-like aperture array can reject the other inputs. Furthermore, by designing a metasurface with multiple OAM-selective holographic phase profiles superimposed, multitudes of images can be embedded with each image uniquely activated by the corresponding input beam with the correct OAM. This provides a systematic way of OAM-"multiplexed" holography.

Building on these principles, we extended the concept to TAM vectorial holography by exploiting the polarization-multiplexing and manipulation capabilities of bi-layer metasurfaces. Figure 2 presents an overview of the proposed TAM vectorial holography device, which generates independent sets of two-dimensional point-like target holographic images according to the TAM input states. In the schematics in Figure 2(a), each TAM input state $|p, l\rangle$ can be defined by its polarization $p$ and helical mode index $l$, representing SAM and OAM, respectively. As shown in the right columns of Figure 2(a), the regions masked by the aperture array are depicted in black throughout this work. The bi-layer metasurface, with its unit cell schematics and scanning electron microscopy (SEM) images shown in Figure 2(b) and (c), transforms each TAM state into the desired vectorial holographic images. Three target images per output channel are presented to visualize their independent designability. These images correspond to the far-field intensity and polarization profiles, with the latter represented by two spherical coordinates (azimuth and elevation angles) of polarization on the Poincaré sphere (Figure 2(d)). For multiplexing, orthogonal TAM input states were considered; two TAM input states are orthogonal ($\langle p_1, l_1 | p_2, l_2\rangle = 0$) when their polarizations are orthogonal ($\langle p_1 | p_2\rangle = 0$) or their helical mode indices differ ($l_1 \neq l_2$). Notably, our scheme accommodates arbitrary elliptical polarization states as input polarizations.

Recent advancements in metasurfaces have widened the controllability of the amplitude, phase, and polarization state of light. In particular, bi-layer dielectric metasurfaces have been demonstrated to achieve complete linear control over coherent light transmission [37-39]. Based on the Jones calculus [40], we represented our bi-layer metasurfaces as unitary Jones matrices, where unitarity simplifies the design process and guarantees the maximal power efficiency of the device. Various factors such as material absorption and fabrication imperfections may result in non-unitarity and a decrease in efficiency. Unitary matrices can be further decomposed using four design parameters $\theta_1$, $\theta_2$, $\phi_{1M}$, and $\phi_{1m}$ given as [37]:

$$U = \begin{bmatrix} \cos(2\theta_2 - \theta_1) & \sin(2\theta_2 - \theta_1) \\ \sin(2\theta_2 - \theta_1) & \cos(2\theta_2 - \theta_1) \end{bmatrix} \begin{bmatrix} e^{i\phi_{1M}} & 0 \\ 0 & e^{i\phi_{1m}} \end{bmatrix} \begin{bmatrix} \cos\theta_1 & \sin\theta_1 \\ -\sin\theta_1 & \cos\theta_1 \end{bmatrix}. \quad (1)$$

These design parameters are closely related to the structural parameters of the bi-layer unitcell in Figure 2(b); $\theta_i$ corresponds to the orientation angles of the major axes of the elliptical posts of the bottom ($i = 1$) and top ($i = 2$) layer, and $\phi_{1M,1m}$ are the phase delays of the bottom layer along the major and minor axes, which are directly related to the lengths of the major and minor axes $l_{1M,1m}$ of the bottom layer. The lengths of the major and minor axes $l_{2M,2m}$ of the top layer were fixed over the entire metasurface so that all the nanoposts on the top layer acted as local half-wave retarders with spatially varying rotation angles (Text S1, Supporting Information).

Based on this description, the metasurfaces were optimized using a gradient descent algorithm [41, 42]. Figure 3 represents a flowchart of the optimization algorithm. Given a TAM input state and a set of target images discretized by a two-dimensional sampling array, random initial parameters are iteratively updated by minimizing the difference between the calculated far-field profiles and the targets. During optimization, partial derivatives are analytically computed using the chain rule (Text S2, Supporting Information). The method extends naturally to multi-objective designs by summing the losses and gradients for each input–target pair.

Figure 4 summarizes the numerical and experimental demonstrations of the proposed TAM vectorial holography. For the demonstrations, we designed a metasurface that produces 12 scalar (four vectorial) holographic images in response to four different TAM input states. Figure 4(a) depicts the chosen four TAM states with the polarization states of $|p_1\rangle = [1; 0]$, $|p_{1,\perp}\rangle = [0; 1]$, $|p_2\rangle = [\cos\frac{\pi}{3}; e^{i\frac{\pi}{2}} \sin\frac{\pi}{3}]$, and $|p_{2,\perp}\rangle = [-\sin\frac{\pi}{3}; e^{i\frac{\pi}{2}} \cos\frac{\pi}{3}]$ and the helical mode indices of $l_1$=-2 (for $|p_1\rangle$ and $|p_{1,\perp}\rangle$) and $l_2$=2 (for $|p_2\rangle$ and $|p_{2,\perp}\rangle$), respectively. Each TAM input state was intended to generate different sets of three scalar images in its intensity and polarization distributions: Latin letters (“A,” “B,” and “C”), Greek letters (“α,” “β,” and “γ”), numbers (“1,” “2,” and “3”), and animal icons (dog, cat, and duck), respectively (refer to Figure S1, Supporting Information for the ground truth images). In practice, these three scalar images for each TAM input would be designed such that a single fully vectorial target holographic image with coupled intensity and polarization distributions is formed as a result. However, in such cases, a high degree of correlation typically exists between the three images. Thus, to demonstrate their independent designability even more clearly, we specifically chose these examples such that each scalar image was completely different from the others. For easy visual inspection of the reproduction quality, target images were binarized for intensity ($I$=0.67, 0.33), azimuth ($\psi$=±π/4), and elevation ($\chi$=±π/8). However, the design method itself supports arbitrary, non-binarized target values. To apply OAM multiplexing, a two-dimensional sampling function was used to discretize the target images in a square lattice with a period of 0.08 in the normalized spatial-frequency domain. For clear visualization, the output channels for different helical mode indices were spatially separated (refer to Text S3, Supporting Information for the detailed configuration).

Figure 4(b) presents the numerical reconstruction of the holographic images, which clearly reproduce the intended intensity and polarization distributions for each TAM input state. The agreement between the reconstructed and target images was evaluated across all 12 images (intensity and polarization), yielding an average correlation coefficient of ≈0.962. Polarization fidelity was further assessed from the cosine similarity of the Stokes parameter distributions of the four output channels, giving an average value of ≈0.963. These results confirmed the high fidelity of the numerical design (refer to Experimental Section for definitions of these measures). Additional discussion regarding channel crosstalk is provided in Text S4, Supporting Information. The optimized efficiencies of the four output channels, defined as the ratio of the power passing through the aperture array within the target region to the incident power, were ≈19.5%, 16.7%, 16.6%, and 18.9%, respectively, assuming unitary transmission. Note that the maximum efficiency, which is determined by the number of encoded helical modes in the OAM-multiplexing mechanism, is generally considerably lower than unity.

Figure 4(c) shows the experimental reconstruction of the holographic images. Details of the post-processing procedure for the measured images can be found in Text S5, Supporting Information. The optimized design was fabricated using well-established electron beam lithography and reactive-ion etching, and was optically characterized using a Stokes parameter measurement setup (Experimental Section). For these measurements, an input-generating metasurface was positioned directly in front of the hologram. A detailed description of these input-generating metasurfaces and the optical characterization configurations can be found in Text S6, Supporting Information. The close match between the simulation and experiment (Figures 4(b) and (c)) clearly demonstrates the feasibility of the proposed TAM vectorial holography. This proves that a fully vectorial target holographic image can be realized for each TAM input state even when some of the input states share the same OAM and differ only in their polarizations, which is impossible with previous single-layer metasurfaces.

## 2.2. Vector beam-multiplexed vectorial holography

When multiple TAM states are collinearly superposed, the resulting field forms a generalized vector beam (VB) with complex polarization and phase distributions. The simplest case is the superposition of two TAM states. Such a vector beam $|\psi\rangle$ can be represented as $|\psi\rangle = \alpha|p_1, l_1\rangle + \beta|p_2, l_2\rangle$, where $\alpha$ and $\beta$ are the complex amplitudes of the constituent states. The detailed beam profile is determined based on their relative magnitudes and phase differences.

A higher-order Poincaré sphere (HOPS) offers a concise visualization of all possible VBs [43]. In the HOPS representation (Figure 5(a)), the two constituent TAM states with potentially different OAMs ($l_1 \neq l_2$) correspond to the north and south poles. This generalizes the conventional Poincaré sphere, in which the north and south poles are usually circular polarization states, and no explicit consideration for OAM is given. Similar to the conventional Poincaré sphere, the elevation and azimuth represent the relative magnitudes and phase differences of the two TAM states corresponding to both poles. Thus, every point on the sphere corresponds to a unique VB state.

The use of generalized VBs instead of simple TAM states extends the input degrees of freedom in optical holography [33-35]. The proposed VB-multiplexed vectorial holography generates a target vectorial hologram only when the correct VB is incident on the device. Previous VB-based multiplexing approaches were limited in polarization diversity and only one VB state per HOPS with different polarizations and helical mode indices could be employed for multiplexing vectorial holograms due to the lack of controllability of light transmission [36]. By contrast, we rigorously validated the extended capability of our approach by deliberately choosing VBs that were as similar to one another as possible without violating orthogonality, thereby demonstrating the independent designability of the resulting vectorial holograms even in such challenging cases. As a result, our design enables the maximal number of two independent input states to be selected from a single HOPS, so that four VBs can be utilized as input states from just two HOPSs instead of four. Furthermore, we put additional constraints on the two HOPSs so that they are as closely related to each other as possible: their polarization states on the north and south poles are identical between HOPSs and only their helical mode numbers were flipped (i.e., the north and south poles have $l_1$ and $l_2$ in the first HOPS, and $l_2$ and $l_1$ in the second). We selected a pair of antipodal points in each HOPS as the input VB states (Figure 5(a)). Although we used two TAM states to form each VB, this concept can be readily extended to more complex VBs composed of three or more TAM states.

Figure 5 shows the experimental realization of VB-multiplexed vectorial holography. We designed and fabricated a metasurface sample based on the same procedures described in the previous section but with VB input states. The constituent TAM states had polarization $|p\rangle = [\cos\frac{\pi}{3}; e^{i\frac{\pi}{2}} \sin\frac{\pi}{3}]$ and helical mode indices $l_1$=-1, $l_2$=2. The first HOPS comprised $|p, l_1\rangle$ and $|p_\perp, l_2\rangle$, and the second $|p, l_2\rangle$ and $|p_\perp, l_1\rangle$ (Figure 5(a)). Without loss of generality, we chose two VB states from each HOPS (i.e., four VB states from four the TAM states), marked by red dots on each HOPS in Figure 5(a). Each input VB was intended to generate different categories of scalar images, in a similar manner with the previous section: Latin letters ("D," "E," and "F"), Greek letters ("δ," "ε," and "ζ"), numbers ("4," "5," and "6"), and weather icons (sun, moon, and cloud), respectively (refer to Figure S1, Supporting Information for the ground truth images). Figure 5(b) shows the polarization and phase distributions of the chosen VBs. As the constituent TAM states were assumed to have uniform intensity distributions and orthogonal polarizations, the resulting VBs exhibited uniform intensity distributions with radially constant polarization and phase distributions.

To prepare the desired VB inputs, we fabricated input-generating metasurfaces (Text S6, Supporting Information). Figure 5(c) presents the post-processed experimental results, which clearly reconstruct the target intensity and polarization images. These results agree well with the numerical predictions (Figure S6, Supporting Information), confirming the effectiveness of the proposed VB-multiplexed vectorial holography. Note that the additional multiplexing dimensions provided by this approach can significantly enhance the versatility of metasurface-based holography and enable new possibilities in applications such as new and more secure encryption schemes in optical communication [41, 44, 45].

### 2.3. Bidirectional TAM vectorial holography

The bidirectional asymmetric transmission of light, where an optical system produces distinct optical responses depending on the illumination direction, holds considerable potential for expanding optical functionalities [40, 46-48]. Our recent work demonstrated that bi-layer metasurfaces can realize polarization–direction-multiplexed vectorial holograms, in which different vectorial holographic images are generated depending on both the illumination direction and polarization state [42]. Despite the potential benefits of bidirectionality, its combination with OAM holography has rarely been explored (with one exception in the field of nonlinear OAM holography [49]). In this section, we propose bidirectional TAM vectorial holography that seamlessly integrates bidirectionality with TAM vectorial holography to add a new dimension to the multiplexing capability of metasurface-based holography.

Figure 6 presents conceptual schematics and experimental results of the proposed approach. To obtain independent TAM–direction-multiplexed vectorial holograms, the transmission space was spatially partitioned, with the lower half-space blocked or discarded (Figure 6(a)) [42]. Using the proposed design method, with a minor adaptation that simultaneously considers illumination from both directions, we designed and fabricated a metasurface sample to demonstrate bidirectional TAM vectorial holography. For front-side illumination, the four chosen TAM states were $|p_1, l_1\rangle$, $|p_{1,\perp}, l_1\rangle$, $|p_2, l_2\rangle$, and $|p_{2,\perp}, l_2\rangle$, where $|p_1\rangle = [1; 0]$, $|p_2\rangle = \frac{1}{\sqrt{2}}[1; \mathrm{i}]$, $l_1$=-1, and $l_2$=3. For back-side illumination, the chosen TAM states were $|p_3, l_1\rangle$, $|p_{3,\perp}, l_1\rangle$, $|p_4, l_2\rangle$, and $|p_{4,\perp}, l_2\rangle$ where $|p_3\rangle = [\cos\frac{\pi}{3}; \sin\frac{\pi}{3}]$, $|p_4\rangle = \frac{1}{\sqrt{2}}[1; 1]$. Each TAM input state was designed to generate different categories of images: Latin letters ("A," "B," and "C"), Greek letters ("α," "β," and "γ"), numbers ("1," "2," and "3"), and animal icons (dog, cat, and duck) for front-side illumination and Latin letters ("D," "E," and "F"), Greek letters ("δ," "ε," and "ζ"), numbers ("4," "5," and "6"), and weather icons (sun, moon, and cloud) for back-side illumination. In total, 24 distinctive intensity or polarization images were engraved on a single device.

Figure 6(b) shows the experimentally reconstructed images. The post-processed results clearly reproduced the intended target intensity and polarization images in good agreement with the numerical predictions (Figure S7, Supporting Information). This demonstrates for the first time that it is possible to integrate bidirectionality and OAM holography in a linear optics platform and that the principle can be readily integrated into various optical systems as it does not rely on nonlinearity.

## 3. Conclusion

We proposed a metasurface-based platform for total angular momentum-multiplexed vectorial holography, enabling the generation of independent vectorial holographic images depending on both the spin and orbital angular momenta—equivalently, the TAM state—of the incident light. For a given helical mode, conventional OAM holography has been limited to generating either intensity-only images for two orthogonal input polarizations or vectorial images for a single input polarization, due to the inherent lack of chirality control in single-layer metasurfaces. In contrast, our approach based on bi-layer metasurfaces, which are capable of arbitrary control over anisotropy and chirality, overcomes this limitation and enables the generation of independent vectorial holographic images for each orthogonal TAM input state. The feasibility of the proposed TAM vectorial holography was validated both numerically and experimentally, demonstrating clear advantages over existing single-layer OAM holography methods. Furthermore, we extended the concept by integrating it with recent advances in vector beam design and bidirectional asymmetric transmission, introducing additional degrees of freedom for input states such as complex coefficients defining vector beams and the direction of illumination.

Looking ahead, the proposed platform can be further improved in three key directions: input states, output functionalities, and metasurface device architectures. From the input perspective, while this work focused on TAM states defined by simple helical modes with integer indices, future implementations could incorporate more complex phase structures beyond conventional OAM modes to represent high-dimensional structured beams [50-57]. Multi-wavelength [58-61] or multi-incidence-angle [28, 29] operations could also enable more diverse forms of multiplexed vectorial holography, further enhancing the information capacity. From the output perspective, although this work focused on far-field vectorial holography, recent developments in three-dimensional OAM holography [49, 60, 62] suggest that extending our platform to volumetric information could unlock new functionalities in optical data storage, displays, and optical encryption. Finally, from the device perspective, integration with nonlinear metasurfaces could provide intensity-dependent or frequency-mixing functionalities, adding new dimensions of light–matter interactions beyond the linear regime [49, 63, 64]. In parallel, active metasurface platforms can introduce dynamic tunability, allowing the real-time reconfiguration of holographic images in response to external stimuli [65-68]. In addition, cascading multiple metasurfaces may enable more complex information processing or enhance multiplexing capacity by sequentially manipulating different optical degrees of freedom [69, 70]. Overall, we envision that the TAM vectorial holography platform presented here will serve as a foundation for next-generation multi-dimensional optical systems.

## 4. Experimental Section

### *Fabrication of Bi-Layer Metasurfaces*

The fabrication began with the deposition of a 790 nm-thick amorphous Si layer on a quartz substrate using plasma-enhanced chemical vapor deposition (PECVD). For electron-beam lithography (EBL), a multilayer stack comprising an adhesion promoter (AR 300-80, Allresist), a resist (AR-P 6200.04, Allresist), and a conductive polymer (AR-PC 5090.02, Allresist) was applied by spin coating. Alignment keys for registering the top and bottom layers were patterned through photolithography, followed by electron-beam (e-beam) evaporation to deposit 10 nm-thick Cr and 100 nm-thick Au, and a subsequent lift-off process. The nanoposts in the bottom layer were patterned by EBL. A 60 nm-thick alumina hard mask was patterned using e-beam evaporation and a lift-off process. Deep reactive ion etching (RIE) was performed to obtain high-aspect-ratio structures with nearly vertical sidewalls. The entire bottom layer was subsequently encapsulated with a 1450 nm-thick SU-8 layer via spin coating, serving both as a protective layer and as a planarized base for the top layer. The top 650 nm-thick amorphous Si layer was then deposited via radio-frequency sputtering, and the top layer nanoposts were patterned using a fabrication process identical to that of the bottom layer.

### *Design of Bi-Layer Metasurfaces*

The bi-layer metasurfaces, which were designed to operate at a wavelength of 915 nm, consisted of a $SiO_2$ substrate, Si nanoposts, and an SU-8 spacer layer (Figure 2(b)). Each unit cell has a lateral period of 450 nm. The heights of the lower and upper Si nanoposts were 790 and 650 nm, respectively, and the SU-8 spacer thickness was 1450 nm. To ensure fabrication compatibility, the major and minor axes of the nanoposts were constrained between 100–380 nm. Within this dimensional range, the optical response was simulated using finite-difference time-domain (FDTD) software provided by Ansys Lumerical Inc. These simulations were used to build a structural library for designing the TAM vectorial holograms. The complex refractive indices used in the simulations were 1.45 ($SiO_2$), 3.61 + 0.0066i (PECVD-grown Si), 3.7 + 0.05i (sputter-grown Si), and 1.56 (SU-8).

***Design of TAM vectorial holography***

The required distributions of parameters ($\phi_{1M}$, $\phi_{1m}$, $\theta_1$, and $\theta_2$) for holograms were optimized using a gradient descent method. For optimization, the loss function for the $i$-th incidence TAM condition was defined as

$$L^{(i)} = \sum_{p,q} \left\{ \left( \left| E_{pq,x}^{(i)} \right|^2 - \left| t_{pq,x}^{(i)} \right|^2 \right)^2 + \left( \left| E_{pq,y}^{(i)} \right|^2 - \left| t_{pq,y}^{(i)} \right|^2 \right)^2 + \left| E_{pq,y}^{(i)} t_{pq,x}^{(i)} - E_{pq,x}^{(i)} t_{pq,y}^{(i)} \right|^2 \right\} \quad (2)$$

where the subscripts $p$ and $q$ represent the indices in the spatial-frequency ($k_x$ and $k_y$) domain, $E_x^{(i)}$ and $E_y^{(i)}$ are the $x$- and $y$-components of the output far-field electric field, respectively, and $t_x^{(i)}$ and $t_y^{(i)}$ represent the corresponding target field components. The first and second terms ensure intentisy matching of the $x$- and $y$-polarization components, whereas the third term enforces relative phase matching. The total loss function is given by $L^{\text{tot}} = \sum_i L^{(i)}$. The gradient of $L^{\text{tot}}$ with respect to the design parameters was obtained using the chain rule, and the parameters were iteratively updated to minimize $L^{\text{tot}}$. Repeating this process yielded optimal parameter distributions that generated the desired holographic images (refer to Text S2, Supporting Information).

***Metrics for Evaluating Fidelity in TAM Holography***

To evaluate the fidelity of the numerical reconstruction from the optimized design, we employed two complementary metrics: the correlation coefficient, for assessing image similarity and the cosine similarity of the Stokes parameter distributions, for assessing polarization fidelity.

The correlation coefficient (CC) between the reconstructed image I and its corresponding target image T, was defined as

$$\text{CC} = \frac{\sum_{a,b} \left( I_{a,b} - \bar{I} \right) \left( T_{a,b} - \bar{T} \right)}{\sqrt{\sum_{a,b} \left( I_{a,b} - \bar{I} \right)^2} \sqrt{\sum_{a,b} \left( T_{a,b} - \bar{T} \right)^2}} \quad (3)$$

where $I_{a,b}$ and $T_{a,b}$ denote the intensities of the ($a$, $b$)-th pixel in the reconstructed and target images, respectively, and $\bar{I}$ and $\bar{T}$ are their mean values. This measure was applied across all 12 holographic images (four intensity and eight polarization-related images).

For polarization fidelity, we computed the cosine similarity (CS) between the reconstructed and target Stokes parameter images, which is defined as

$$\mathrm{CS} = \frac{\overrightarrow{S_{\mathrm{r}}} \cdot \overrightarrow{S_{\mathrm{t}}}}{\left\|\overrightarrow{S_{\mathrm{r}}}\right\| \left\|\overrightarrow{S_{\mathrm{t}}}\right\|} \tag{4}$$

where $\overrightarrow{S_{\mathrm{r}}}$ and $\overrightarrow{S_{\mathrm{t}}}$ denotes the reconstructed and target Stokes parameter vectors, which can be derived from the azimuth and elevation images for each output channel.

***Optical Characterization***

The fabricated metasurfaces were optically characterized using the setup shown in Figure S5. The polarization state of the incident light was controlled using half-wave and quarter-wave plates. After transmission through the metasurface, a Fourier plane was formed at the back focal plane (BFL) of the objective lens (MPLFLN 10x, Olympus). A lens (LSB04, Thorlabs) with a focal length of 150 mm was used to project the Fourier plane onto the CCD camera (CS505MU, Thorlabs). TAM vectorial holograms were analyzed using the Stokes parameters (Text S7, Supporting Information), which were derived from four or six intensity measurements under specific polarization configurations, involving a linear polarizer and a quarter-wave plate placed in front of the CCD camera [71].

**Author contribution:** J.J. conceived the idea, and J.S. supervised the project. J.J. conducted the theoretical analyses. J.J. and H.K. performed the numerical simulations and designed the samples. J.J. fabricated the samples and characterized them optically. J.J., H.K., and J.S. prepared the manuscript. All authors have accepted responsibility for the entire content of this submitted manuscript and approved its submission.

**Acknowledgments:** This work is supported by National Research Foundation (NRF) grants (RS-2024-00414119, RS-2021-NR057359) funded by the Ministry of Science and ICT (MSIT), Republic of Korea.

**Data availability statement:** The data that support the findings of this study are available from the corresponding author upon reasonable request.

**Conflict of interest statement:** The authors declare no conflicts of interest.

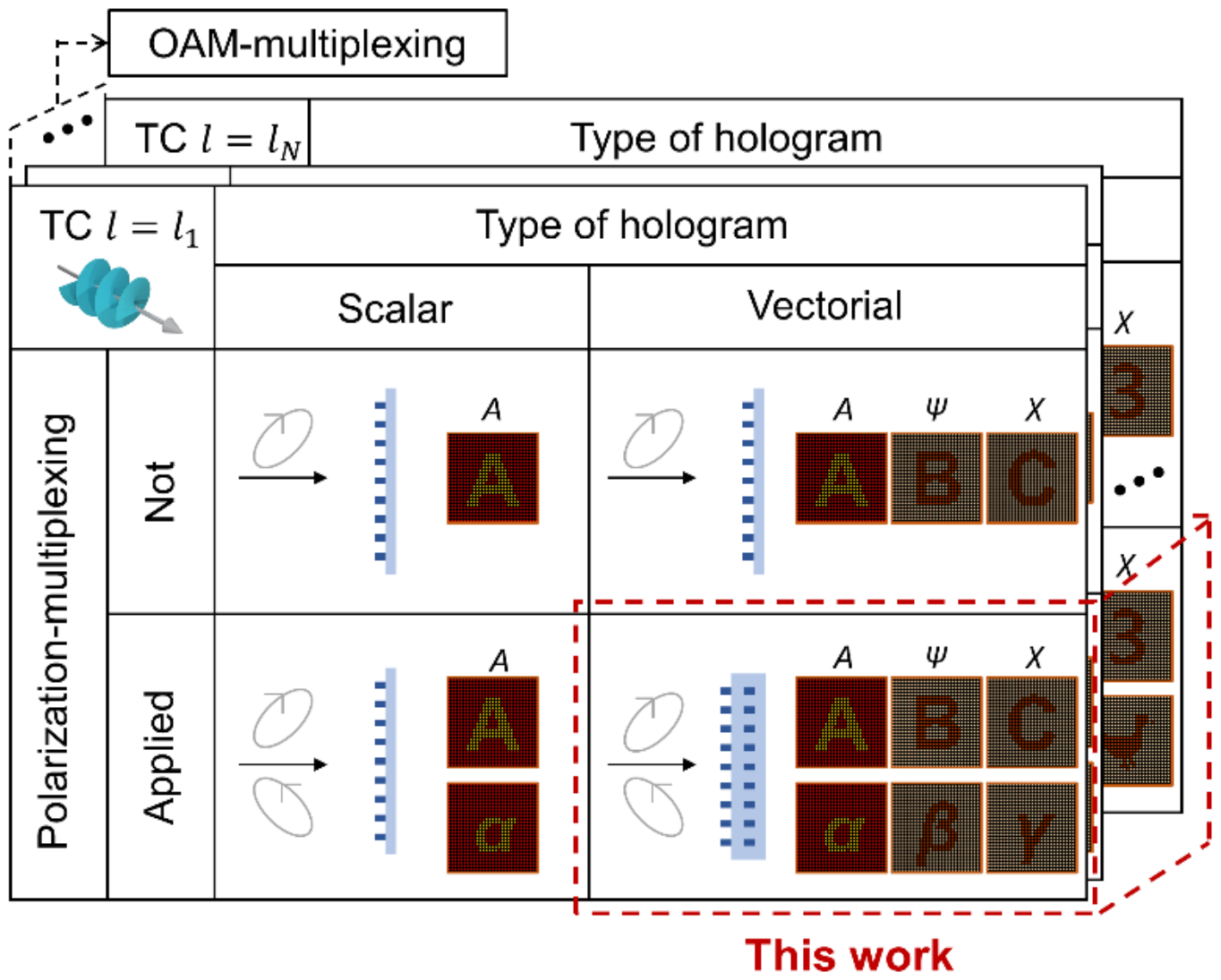

**Figure 1. Classification of OAM holography.** Comparison OAM holography in terms of hologram type (scalar or vectorial) and polarization-multiplexing capability. The proposed approach achieves vectorial holography with polarization–OAM multiplexing.

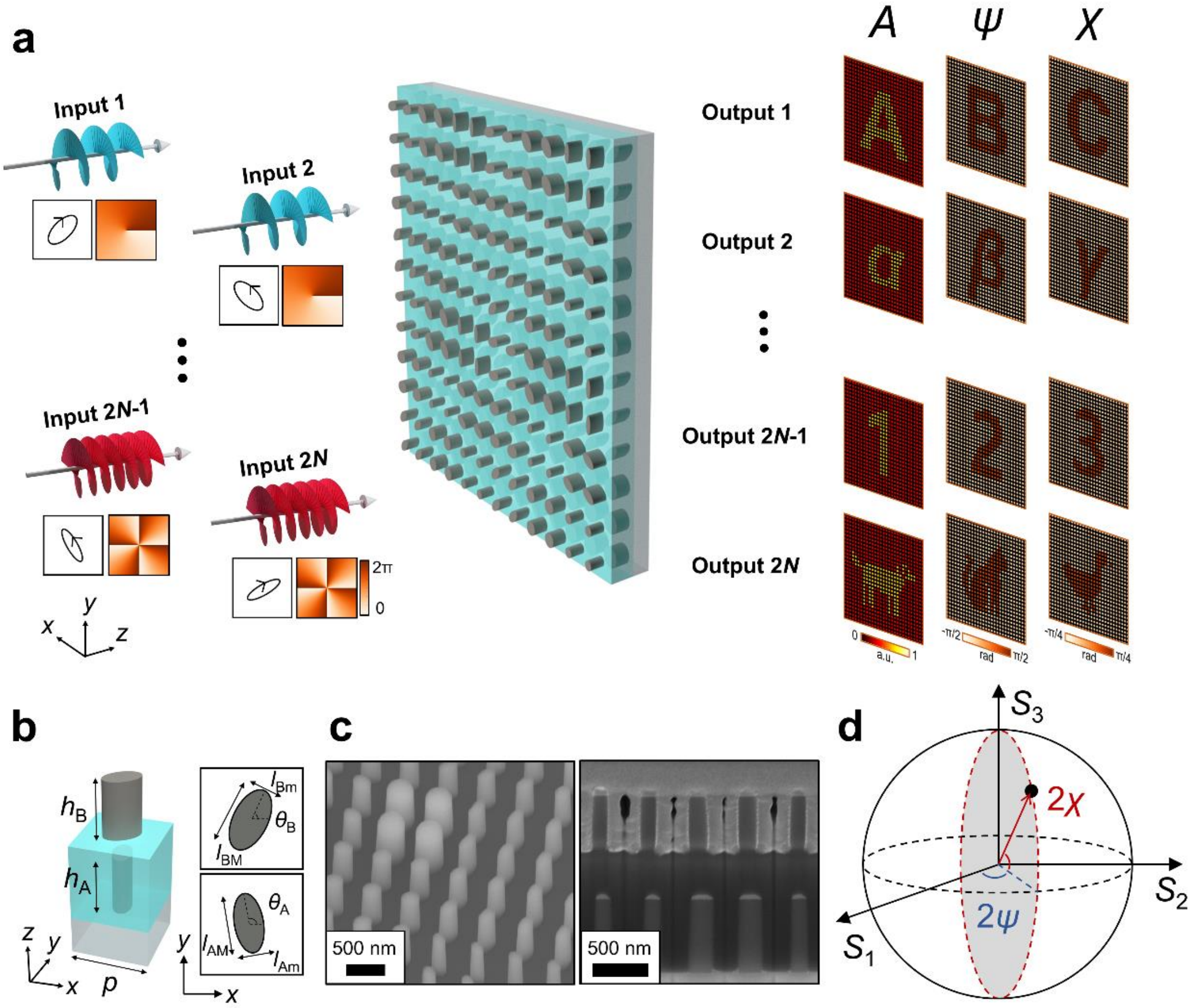

**Figure 2. TAM vectorial holography.** (a) Schematic of the TAM vectorial holography concept. Polarizations and phases, defining input TAM states, are represented in the square box. Three target images for each output channel are encoded using the intensity (A), azimuth ($\psi$), and elevation ($\chi$). (b) Unit cell structure of the bi-layer metasurface. The right panels show cross sections of the top (the subscript B) and bottom (the subscript A) layers of the bi-layer structure. $p$: period; $h$: height of posts; $l_{\mathrm{M,m}}$: length of the major (M) and minor (m) axes; $\theta$: orientation angle. (c) Scanning electron microscopy images of the fabricated bi-layer metasurface. (d) Representation of azimuth and elevation for image encoding on the Poincaré sphere.

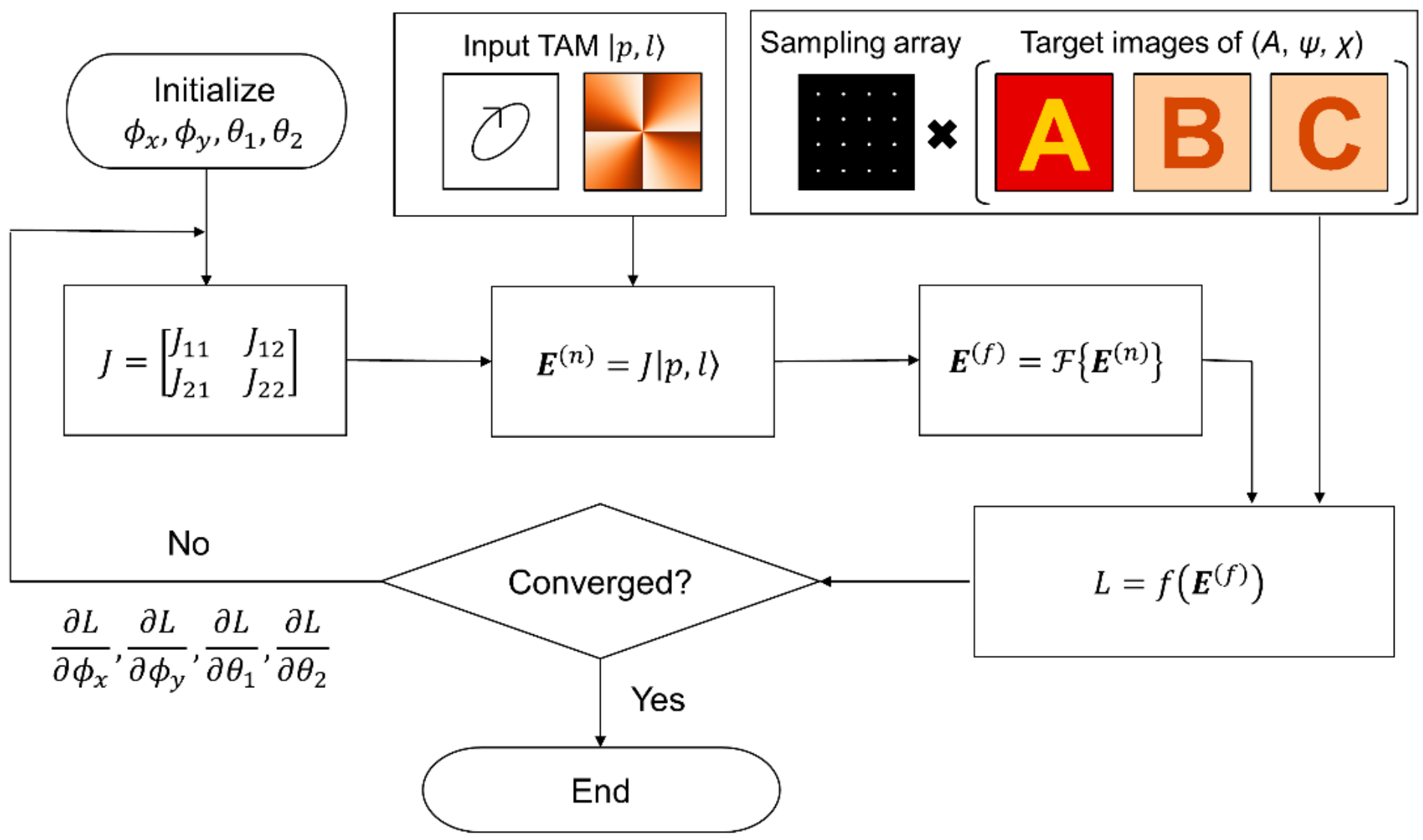

**Figure 3. Optimization workflow for TAM vectorial holography.** Flowchart of the gradient descent optimization used to optimize bi-layer metasurface parameters.

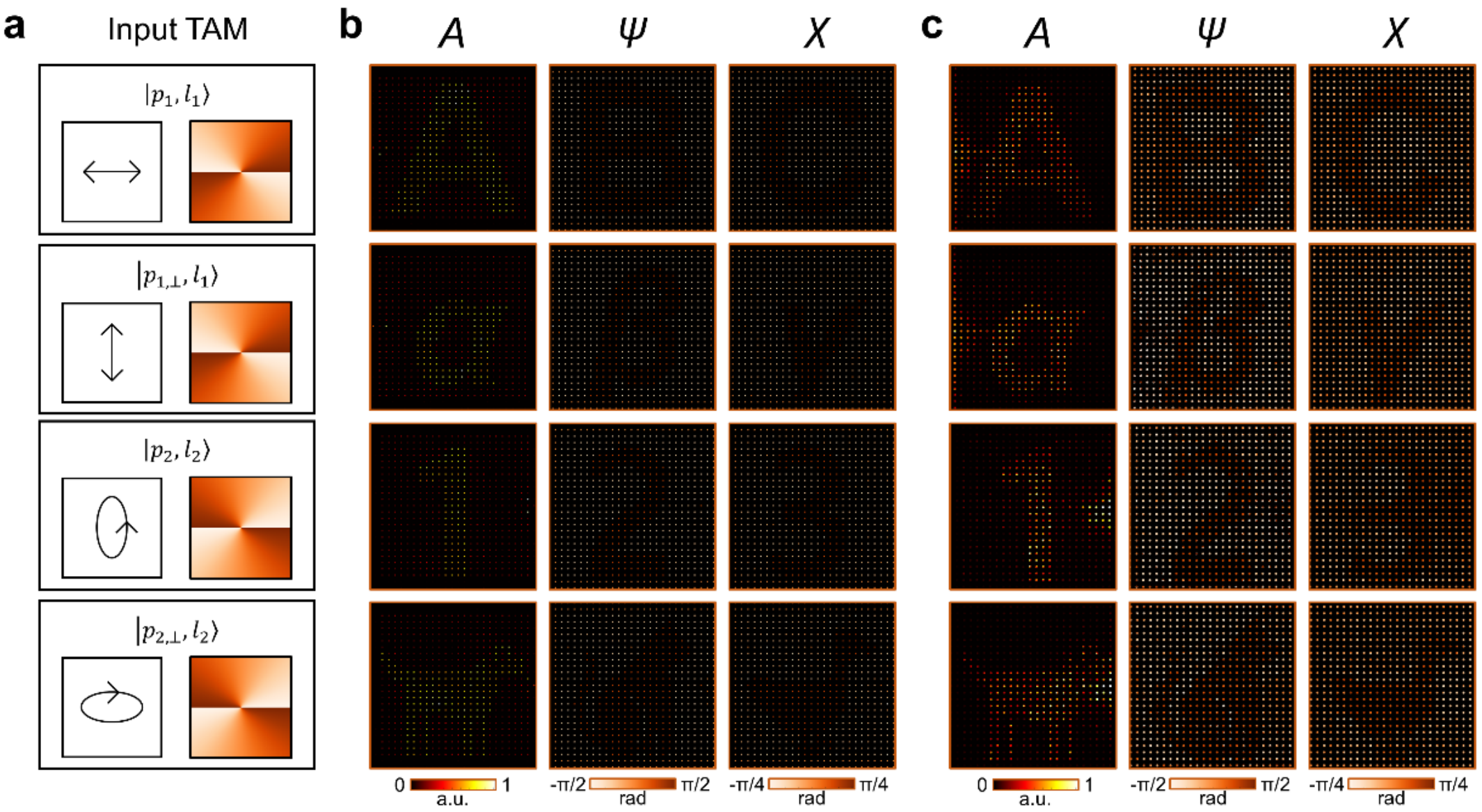

**Figure 4. Numerical and experimental demonstrations of TAM vectorial holography.** (a) Polarizations and phases of four input TAM states. (b) Numerically reconstructed holographic images. (c) Experimentally reconstructed holographic images.

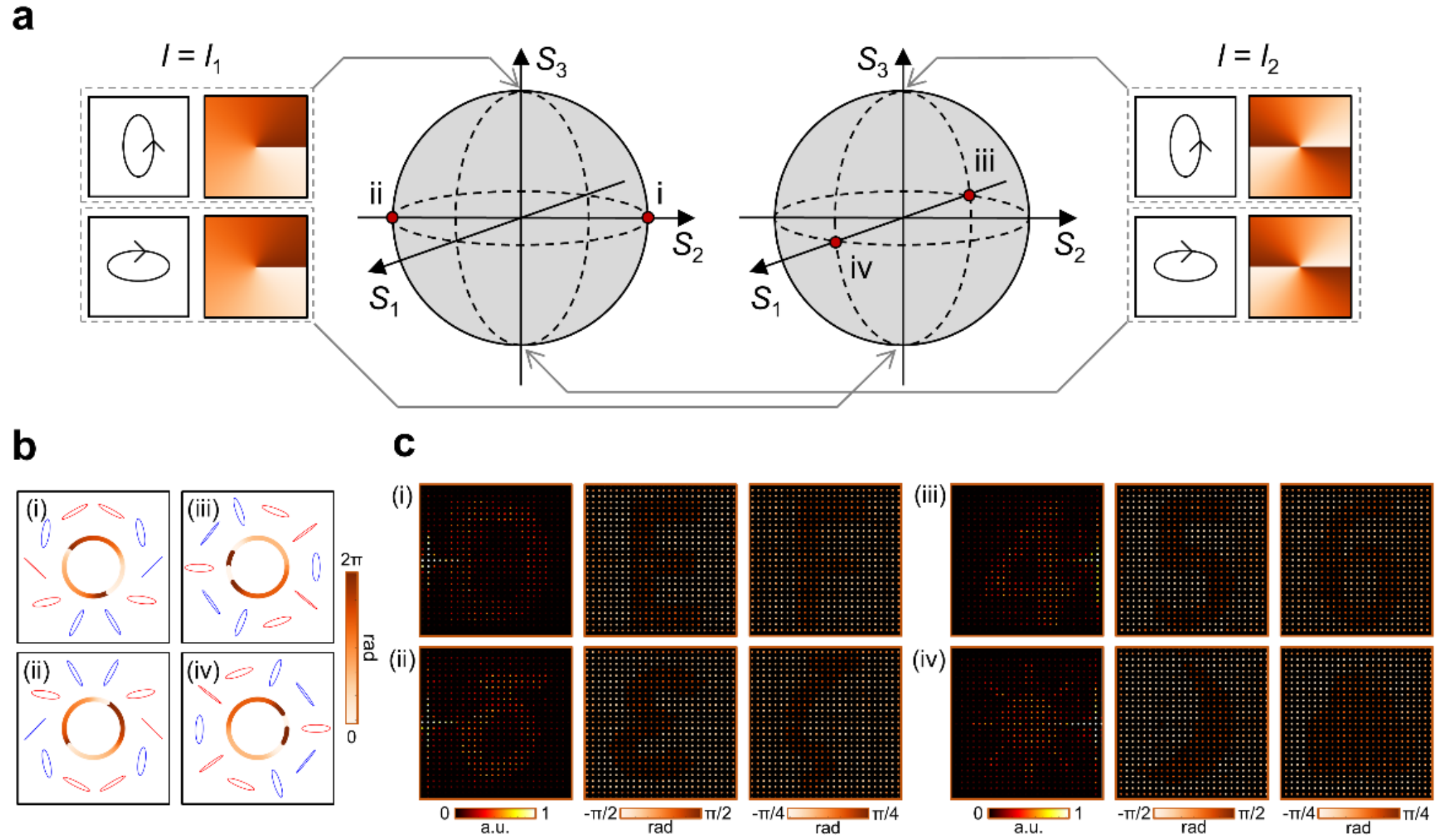

**Figure 5. Vector beam-multiplexed TAM vectorial holography.** (a) Higher-order Poincaré spheres representing the vector beam (VB) space. Each sphere is defined by two TAM states with a pair of orthogonal polarizations $p$ and $p_\perp$ and different helical mode indices $l_1$ and $l_2$. Red dots indicate the selected input VBs. (b) Polarization and phase profiles of the selected input VBs. (c) Experimental reconstructions of VB-multiplexed vectorial holography.

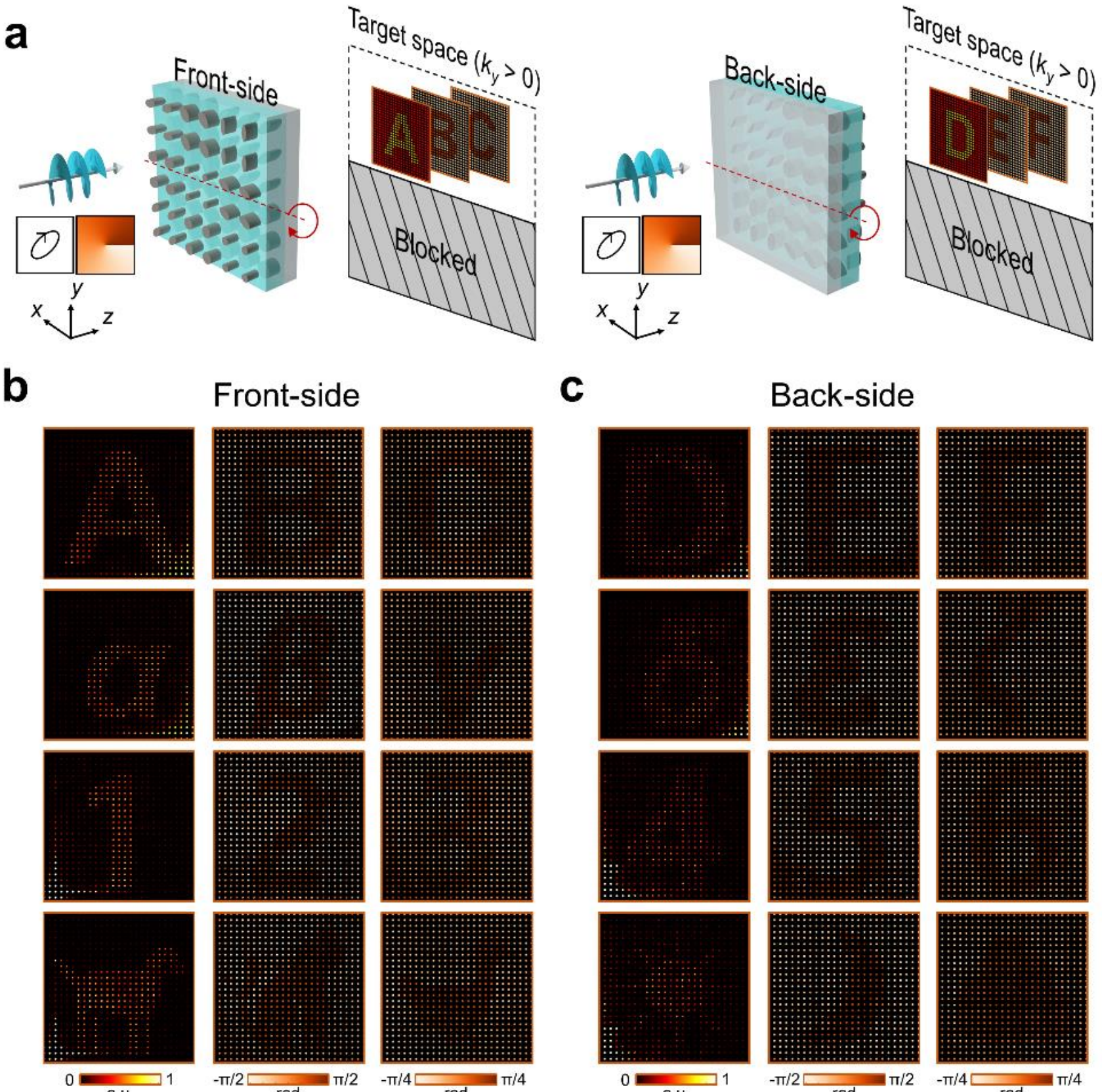

**Figure 6. Bidirectional TAM vectorial holography.** (a) Conceptual schematic of bidirectional operation, where transmission space is partitioned and the lower half-space is blocked. The metasurface sample is rotated about the *x*-axis to switch the direction of illumination. Experimental reconstructions for (b) front-side illumination and (c) back-side illumination, demonstrate the TAM–direction-multiplexing capability of bidirectional TAM vectorial holography.

## References

1. Gabor, D., *A new microscopic principle.* 1948.

2. Leith, E.N. and J. Upatnieks, *Reconstructed wavefronts and communication theory.* Journal of the optical society of America, 1962. **52**(10): p. 1123-1130.

3. Schnars, U., et al., *Digital holography and wavefront sensing.* Digital Holography, 2015.

4. Márquez, A., et al., *Information multiplexing from optical holography to multi-channel metaholography.* Nanophotonics, 2023. **12**(24): p. 4415-4440.

5. Jung, J., et al., *Broadband metamaterials and metasurfaces: a review from the perspectives of materials and devices.* Nanophotonics, 2020. **9**(10): p. 3165-3196.

6. Huang, L., S. Zhang, and T. Zentgraf, *Metasurface holography: from fundamentals to applications.* Nanophotonics, 2018. **7**(6): p. 1169-1190.

7. Zou, Y., et al., *Metasurface holography with multiplexing and reconfigurability.* Nanomaterials, 2023. **14**(1): p. 66.

8. Jeong, H.-D., H. Kim, and S.-Y. Lee, *Review of metasurfaces with extraordinary flat optic functionalities.* Current Optics and Photonics, 2024. **8**(1): p. 16-29.

9. Ayoub, A.B. and D. Psaltis, *High speed, complex wavefront shaping using the digital micro-mirror device.* Scientific Reports, 2021. **11**(1): p. 18837.

10. Park, J.-H. and B. Lee, *Holographic techniques for augmented reality and virtual reality near-eye displays.* Light: Advanced Manufacturing, 2022. **3**(1): p. 137-150.

11. Xiong, J., et al., *Holographic optical elements for augmented reality: principles, present status, and future perspectives.* Advanced Photonics Research, 2021. **2**(1): p. 2000049.

12. Pi, D., J. Liu, and Y. Wang, *Review of computer-generated hologram algorithms for color dynamic holographic three-dimensional display.* Light: Science & Applications, 2022. **11**(1): p. 231.

13. Wu, L. and Z. Zhang, *Domain multiplexed computer-generated holography by embedded wavevector filtering algorithm.* PhotoniX, 2021. **2**(1): p. 1.

14. Winzer, P.J. *Modulation and multiplexing in optical communications*. in *Conference on Lasers and Electro-Optics*. 2009. Optica Publishing Group.

15. Khonina, S.N., et al., *Optical multiplexing techniques and their marriage for on-chip and optical fiber communication: a review.* Opto-Electronic Advances, 2022. **5**(8): p. 210127-1-210127-25.

16. Fabre, C. and N. Treps, *Modes and states in quantum optics.* Reviews of Modern Physics, 2020. **92**(3): p. 035005.

17. Luo, W., et al., *Recent progress in quantum photonic chips for quantum communication and internet.* Light: Science & Applications, 2023. **12**(1): p. 175.

18. Hu, J., et al., *Diffractive optical computing in free space.* Nature Communications, 2024. **15**(1): p. 1525.

19. Bai, Y., et al., *Photonic multiplexing techniques for neuromorphic computing.* Nanophotonics, 2023. **12**(5): p. 795-817.

20. Fang, X., H. Ren, and M. Gu, *Orbital angular momentum holography for high-security encryption.* Nature Photonics, 2020. **14**(2): p. 102-108.

21. Ren, H., et al., *Metasurface orbital angular momentum holography.* Nature communications, 2019. **10**(1): p. 2986.

22. Ren, H., et al., *Complex-amplitude metasurface-based orbital angular momentum holography in momentum space.* Nature Nanotechnology, 2020. **15**(11): p. 948-955.

23. Yu, N. and F. Capasso, *Flat optics with designer metasurfaces.* Nature materials, 2014. **13**(2): p. 139-150.

24. Arbabi, A., et al., *Dielectric metasurfaces for complete control of phase and polarization with subwavelength spatial resolution and high transmission.* Nature nanotechnology, 2015. **10**(11): p. 937-943.

25. Dorrah, A.H. and F. Capasso, *Tunable structured light with flat optics.* Science, 2022. **376**(6591): p. eabi6860.

26. Balthasar Mueller, J., et al., *Metasurface polarization optics: independent phase control of arbitrary orthogonal states of polarization.* Physical review letters, 2017. **118**(11): p. 113901.

27. Shi, Z., et al., *Single-layer metasurface with controllable multiwavelength functions.* Nano letters, 2018. **18**(4): p. 2420-2427.

28. Kamali, S.M., et al., *Angle-multiplexed metasurfaces: encoding independent wavefronts in a single metasurface under different illumination angles.* Physical Review X, 2017. **7**(4): p. 041056.

29. Kim, N., et al., *Highly angle-sensitive and efficient optical metasurfaces with broken mirror symmetry.* Nanophotonics, 2023. **12**(13): p. 2347-2358.

30. Zhou, H., et al., *Polarization-encrypted orbital angular momentum multiplexed metasurface holography.* ACS nano, 2020. **14**(5): p. 5553-5559.

31. Yang, H., et al., *Angular momentum holography via a minimalist metasurface for optical nested encryption.* Light: Science & Applications, 2023. **12**(1): p. 79.

32. Wang, Y., et al., *Orbital angular momentum multiplexing holography based on multiple polarization channel metasurface.* Nanophotonics, 2023. **12**(23): p. 4339-4349.

33. Xie, Z., et al., *Cylindrical vector beam multiplexing holography employing spin-decoupled phase modulation metasurface.* Nanophotonics, 2024. **13**(4): p. 529-538.

34. Zhou, C., et al., *Optical vectorial-mode parity Hall effect: a case study with cylindrical vector beams.* Nature Communications, 2024. **15**(1): p. 4022.

35. He, H., et al., *Higher-order Poincaré sphere multiplexed metasurface holography for optical information encryption.* Optics & Laser Technology, 2025. **180**: p. 111555.

36. Yu, Z., et al., *Spin-orbit-locking vectorial metasurface holography.* Advanced Materials, 2025. **37**(9): p. 2415142.

37. Chang, T., et al., *Universal metasurfaces for complete linear control of coherent light transmission.* Advanced Materials, 2022. **34**(44): p. 2204085.

38. Palmieri, A., et al., *Do dielectric bilayer metasurfaces behave as a stack of decoupled single-layer metasurfaces?* Optics Express, 2024. **32**(5): p. 8146-8159.

39. Dorrah, A.H., et al., *Free-standing bilayer metasurfaces in the visible.* Nature Communications, 2025. **16**(1): p. 3126.

40. Menzel, C., C. Rockstuhl, and F. Lederer, *Advanced Jones calculus for the classification of periodic metamaterials.* Physical Review A—Atomic, Molecular, and Optical Physics, 2010. **82**(5): p. 053811.

41. Shi, Z., et al., *Nonseparable polarization wavefront transformation.* Physical Review Letters, 2022. **129**(16): p. 167403.

42. Kim, H., J. Jung, and J. Shin, *Bidirectional Vectorial Holography Using Bi-Layer Metasurfaces and Its Application to Optical Encryption.* Advanced Materials, 2024. **36**(44): p. 2406717.

43. Milione, G., et al., *Higher-order Poincaré sphere, Stokes parameters, and the angular momentum of light.* Physical review letters, 2011. **107**(5): p. 053601.

44. Song, Q., et al., *Vectorial metasurface holography.* Applied Physics Reviews, 2022. **9**(1).

45. Guo, X., et al., *Stokes meta-hologram toward optical cryptography.* Nature Communications, 2022. **13**(1): p. 6687.

46. Fedotov, V., et al., *Asymmetric propagation of electromagnetic waves through a planar chiral structure.* Physical review letters, 2006. **97**(16): p. 167401.

47. Menzel, C., et al., *Asymmetric transmission of linearly polarized light at optical metamaterials.* Physical review letters, 2010. **104**(25): p. 253902.

48. Chen, K., et al., *Directional janus metasurface.* Advanced Materials, 2020. **32**(2): p. 1906352.

49. Shen, F., et al., *3D Orbital Angular Momentum Nonlinear Holography.* Advanced Optical Materials, 2025. **13**(9): p. 2402836.

50. Cheng, P., S. Huang, and C. Yan, *Ellipticity-encrypted orbital angular momentum multiplexed holography.* Journal of the Optical Society of America A, 2021. **38**(12): p. 1875-1883.

51. Zhu, G., et al., *Ultra-dense perfect optical orbital angular momentum multiplexed holography.* Optics Express, 2021. **29**(18): p. 28452-28460.

52. Wang, F., et al., *Angular multiplexation of partial helical phase modes in orbital angular momentum holography.* Optics Express, 2022. **30**(7): p. 11110-11119.

53. Li, F., et al., *Multiple-dimensional multiplexed holography based on modulated chiro-optical fields.* Optics Express, 2022. **30**(23): p. 41567-41579.

54. Zhang, N., et al., *Multiparameter encrypted orbital angular momentum multiplexed holography based on multiramp helicoconical beams.* Advanced Photonics Nexus, 2023. **2**(3): p. 036013-036013.

55. Wang, F., et al., *Enhancing the information capacity with modulated orbital angular momentum holography.* IEEE Photonics Journal, 2022. **14**(1): p. 1-5.

56. Xu, Z., et al., *Multiplexed Holography Using Spiral Fractional Orbital Angular Momentum.* Laser & Photonics Reviews, 2025. **19**(12): p. 2401954.

57. Yuan, H., et al., *Manipulation of Optical Encryption Metasurface Orbital Angular Momentum Holography via Multi-Spatial Modal Basis Multiplexing.* Laser & Photonics Reviews: p. e00661.

58. Jang, J., et al., *Wavelength-multiplexed orbital angular momentum meta-holography.* PhotoniX, 2024. **5**(1): p. 27.

59. He, G., et al., *Multiplexed manipulation of orbital angular momentum and wavelength in metasurfaces based on arbitrary complex-amplitude control.* Light: Science & Applications, 2024. **13**(1): p. 98.

60. Su, H., et al., *Achromatic 3D Multi-Color Orbital Angular Momentum Holography.* Advanced Science, 2025: p. 2503488.

61. Meng, W., et al., *Ultranarrow-linewidth wavelength-vortex metasurface holography.* Science Advances, 2025. **11**(12): p. eadt9159.

62. Kong, L.J., et al., *3D Orbital Angular Momentum Multiplexing Holography with Metasurfaces: Encryption and Dynamic Display of 3D Multi-Targets.* Laser & Photonics Reviews, 2025. **19**(6): p. 2401608.

63. Fang, X., et al., *High-dimensional orbital angular momentum multiplexing nonlinear holography.* Advanced Photonics, 2021. **3**(1): p. 015001.

64. Fang, X., et al., *Multichannel nonlinear holography in a two-dimensional nonlinear photonic crystal.* Physical Review A, 2020. **102**(4): p. 043506.

65. Shaltout, A.M., V.M. Shalaev, and M.L. Brongersma, *Spatiotemporal light control with active metasurfaces.* Science, 2019. **364**(6441): p. eaat3100.

66. Park, J., et al., *Dynamic reflection phase and polarization control in metasurfaces.* Nano letters, 2017. **17**(1): p. 407-413.

67. Kim, J.Y., et al., *Full 2π tunable phase modulation using avoided crossing of resonances.* Nature Communications, 2022. **13**(1): p. 2103.

68. Liao, Y., Y. Fan, and D. Lei, *Thermally tunable binary-phase VO2 metasurfaces for switchable holography and digital encryption.* Nanophotonics, 2024. **13**(7): p. 1109-1117.

69. Jung, J., H. Kim, and J. Shin, *Three-dimensionally reconfigurable focusing of laser by mechanically tunable metalens doublet with built-in holograms for alignment.* Nanophotonics, 2023. **12**(8): p. 1373-1385.

70. Mei, F., et al., *Cascaded metasurfaces for high-purity vortex generation.* Nature Communications, 2023. **14**(1): p. 6410.

71. Berry, H.G., G. Gabrielse, and A. Livingston, *Measurement of the Stokes parameters of light.* Applied optics, 1977. **16**(12): p. 3200-3205.

# Supporting Information for "Arbitrary Total Angular Momentum Vectorial Holography Using Bi-Layer Metasurfaces"

*Joonkyo Jung*[1], *Hyeonhee Kim*[1] and *Jonghwa Shin*[1,*]

[1]Department of Materials Science and Engineering, KAIST, Daejeon 34141, Republic of Korea.

[*]E-mail: qubit@kaist.ac.kr

## Supporting Texts

### S1. Unitary Jones matrix of bi-layer metasurfaces

The Jones calculus is a concise and powerful tool for describing the interaction between light and optical systems [1]. In particular, the seminal work [2] demonstrated that highly transmissive single-layer metasurfaces can be approximated as unitary symmetric Jones matrices. Based on this approximation, we suggested in [3] that by employing bi-layer metasurfaces, the symmetry constraint of Jones matrices can be broken, enabling the realization of arbitrary unitary Jones matrices.

A unitary symmetric matrix can be decomposed as $U_s = R(\theta)\begin{bmatrix} \mathrm{e}^{\mathrm{i}\phi_M} & 0 \\ 0 & \mathrm{e}^{\mathrm{i}\phi_m} \end{bmatrix}R(-\theta)$ where $U_s$, $\phi_M$, $\phi_m$, $R(\theta)$ represent a unitary symmetric matrix, two independent phases, a rotation matrix with angle $\theta$ relative to the *x*-axis. This indicates that a single-layer metasurface supports two linear eigen-polarization states defined by $\theta$, with pure phase responses of $\phi_M$ and $\phi_m$ along the respective eigen-polarizations (that is, major and minor axes).

By cascading two layers of such unitary symmetric matrices, the overall Jones matrix of the system is expressed as

$$U = U_{S,2}U_{S,1} = R(\theta_2)\begin{bmatrix} \mathrm{e}^{\mathrm{i}\phi_{2M}} & 0 \\ 0 & \mathrm{e}^{\mathrm{i}\phi_{2m}} \end{bmatrix}R(-\theta_2)R(\theta_1)\begin{bmatrix} \mathrm{e}^{\mathrm{i}\phi_{1M}} & 0 \\ 0 & \mathrm{e}^{\mathrm{i}\phi_{1m}} \end{bmatrix}R(-\theta_1) \quad \text{(S1)}$$

where $U$ represents a unitary matrix and the subscripts 1 and 2 represent the bottom and top layers as in Figure 2(b) in the main text, respectively. While a unitary matrix intrinsically has four degrees of freedom, our bi-layer metasurfaces possess six degrees of freedom. Due to these residual degrees of freedom, two parameters can be fixed as constants without loss of controllability over the Jones matrices. For simplicity and to reduce structural complexity, we set $\phi_{2M} = 0$ and $\phi_{2m} = \pi$; in other words, the top layer act as a half-wave plate with spatially varying rotation angles. Under this condition, the Jones matrix can be rearranged as

$$U = \begin{bmatrix} \cos(2\theta_2 - \theta_1) & \sin(2\theta_2 - \theta_1) \\ \sin(2\theta_2 - \theta_1) & \cos(2\theta_2 - \theta_1) \end{bmatrix}\begin{bmatrix} e^{i\phi_{1M}} & 0 \\ 0 & e^{i\phi_{1m}} \end{bmatrix}\begin{bmatrix} \cos\theta_1 & \sin\theta_1 \\ -\sin\theta_1 & \cos\theta_1 \end{bmatrix}. \quad \text{(S2)}$$

### S2. Gradient descent optimization

Gradient descent optimization was employed to determine the required distributions of design parameters ($\phi_{1M}$, $\phi_{1m}$, $\theta_1$, and $\theta_2$) for the holograms presented in the main text. If the loss function $L$ is expressed in terms of the far-field electric field component $E^f$ as represented in the experimental section of the main text, the gradient can be computed via the chain rule as follows.

First, the derivatives of $L$ with respect to $E^f$ can be analytically evaluated, yielding $\frac{\partial L}{\partial E^f_{pq,x}}$ and $\frac{\partial L}{\partial E^f_{pq,y}}$, where the subscripts $p$ and $q$ represent indices of the spatial-frequency domain and the subscripts $x$ and $y$ denote the *x*- and *y*-polarization components of the field.

Then, applying the chain rule, the derivatives of the loss function with respect to the output fields at the metasurface plane is given as

$$\frac{\partial L}{\partial E^n_{ab,(x,y)}} = \sum_{p,q} \frac{\partial L}{\partial E^f_{pq,(x,y)}} \frac{\partial E^f_{pq,(x,y)}}{\partial E^n_{ab,(x,y)}} \tag{S3}$$

where $E^n$ is the output field at the metasurface plane and the subscript *a* and *b* represent the indices of the spatial domain.

Using the discrete Fourier transform, the far-field electric field and the output field at the metasurface plane have the relation of $E^f_{pq,(x,y)} = \mathcal{F}\left(E^n_{(x,y)}\right)_{pq} = \sum_{a,b} E^n_{ab,(x,y)}\, \mathrm{e}^{-\mathrm{i}\left(k_x^{(p)} x^{(a)} + k_y^{(q)} y^{(b)}\right)}$ where $\mathcal{F}$ represents the discrete Fourier transform, and the superscripts $a$ and $b(p$ and $q)$ also labels indices of the spatial(spatial-frequency) domain. From this relation, the derivatives of $E^f$ with respect to $E^n$ can be written as $\frac{\partial E^f_{pq,(x,y)}}{\partial E^n_{ab,(x,y)}} = \mathrm{e}^{-\mathrm{i}\left(k_x^{(p)} x^{(a)} + k_y^{(q)} y^{(b)}\right)}$. Substituting this expression into Equation S3, the derivatives of $L$ with respect to $E^n$ is given as

$$\frac{\partial L}{\partial E^n_{ab,(x,y)}} = \sum_{p,q} \frac{\partial L}{\partial E^f_{pq,(x,y)}} \mathrm{e}^{-\mathrm{i}\left(k_x^{(p)} x^{(a)} + k_y^{(q)} y^{(b)}\right)} = N^2 \left[ \mathcal{F}^{-1} \left\{ \left( \frac{\partial L}{\partial E^f_{(x,y)}} \right)^* \right\}_{ab} \right]^* \tag{S4}$$

where *N* is the number of points in the spatial domain, and the superscript * represents the complex conjugate.

Lastly, the analytical form of derivatives of $L$ with respect to $\phi_{1M}$, $\phi_{1m}$, $\theta_1$, and $\theta_2$ can be obtained from the Jones matrix described in Equation S2. If the incident light is given as $E^{\mathrm{in}}_{ab} = \begin{bmatrix} E^{\mathrm{in}}_{ab,x} \\ E^{\mathrm{in}}_{ab,y} \end{bmatrix}$, the output field at the metasurface plane is expressed as $\begin{bmatrix} E^n_{ab,x} \\ E^n_{ab,y} \end{bmatrix} = \begin{bmatrix} U_{ab,11} & U_{ab,12} \\ U_{ab,21} & U_{ab,22} \end{bmatrix} \begin{bmatrix} E^{\mathrm{in}}_{ab,x} \\ E^{\mathrm{in}}_{ab,y} \end{bmatrix} = \begin{bmatrix} U_{ab,11} E^{\mathrm{in}}_{ab,x} + U_{ab,12} E^{\mathrm{in}}_{ab,y} \\ U_{ab,21} E^{\mathrm{in}}_{ab,x} + U_{ab,22} E^{\mathrm{in}}_{ab,y} \end{bmatrix}$. From this expression, the derivatives of $L$ with respect to $\phi_{1M}$, $\phi_{1m}$, $\theta_1$, and $\theta_2$ can be analytically represented as

$$\frac{\partial L}{\partial \phi_{1,ab}} = \sum_{x,y} \left( \frac{\partial L}{\partial E^n_{ab,(x,y)}} \frac{\partial E^n_{ab,(x,y)}}{\partial \phi_{1,ab}} + \frac{\partial L}{\partial E^{n,*}_{ab,(x,y)}} \frac{\partial E^{n,*}_{ab,(x,y)}}{\partial \phi_{1,ab}} \right)$$

$$\frac{\partial L}{\partial \phi_{2,ab}} = \sum_{x,y} \left( \frac{\partial L}{\partial E^{n}_{ab,(x,y)}} \frac{\partial E^{n}_{ab,(x,y)}}{\partial \phi_{2,ab}} + \frac{\partial L}{\partial E^{n,*}_{ab,(x,y)}} \frac{\partial E^{n,*}_{ab,(x,y)}}{\partial \phi_{2,ab}} \right)$$

$$\frac{\partial L}{\partial \theta_{1,ab}} = \sum_{x,y} \left( \frac{\partial L}{\partial E^{n}_{ab,(x,y)}} \frac{\partial E^{n}_{ab,(x,y)}}{\partial \theta_{1,ab}} + \frac{\partial L}{\partial E^{n,*}_{ab,(x,y)}} \frac{\partial E^{n,*}_{ab,(x,y)}}{\partial \theta_{1,ab}} \right)$$

$$\frac{\partial L}{\partial \theta_{2,ab}} = \sum_{x,y} \left( \frac{\partial L}{\partial E^{n}_{ab,(x,y)}} \frac{\partial E^{n}_{ab,(x,y)}}{\partial \theta_{2,ab}} + \frac{\partial L}{\partial E^{n,*}_{ab,(x,y)}} \frac{\partial E^{n,*}_{ab,(x,y)}}{\partial \theta_{2,ab}} \right). \quad \text{(S5)}$$

**S3. Spatial arrangement of different output channels**

As mentioned in the main text, we separated output channels with different helical mode indices in the far-field domain for clear visualization of the reconstructed holographic images. However, they may also be placed at the same position as reported in [4]. Two output channels corresponding to the input modes with the same index but orthogonal polarizations were designed to be spatially overlapped. Figure S2 illustrates the spatial configuration of the output channels for TAM, VB-multiplexed, and bidirectional TAM vectorial holograms. Since each output channel has three target images corresponding to the intensity and polarization distributions, six different target images are located within each spatial-frequency range.

**S4. Discussion on crosstalk**

In an ideal multiplexed system, each orthogonal input mode such as an OAM mode or a polarization state produces an independent response in its own output channel, with zero coupling to other channels. Metasurfaces are well suited for this purpose due to their high degrees of freedom for designable optical responses. However, in practical implementations, residual crosstalk arises due to limitations in the hologram design algorithms as well as fabrication and measurement imperfections. As long as the crosstalk remains sufficiently low, reliable multiplexing can still be achieved. In this section, we discuss two possible sources of crosstalk.

*Crosstalk due to optimization algorithms of metasurface holograms*

For polarization–OAM-multiplexed holography, ideal vectorial far-field distributions can be defined. From these, the ideal Jones matrices are directly calculated from the desired input states and target far-field distributions based on the Fourier transform. While these ideal Jones matrices are composed of four independent complex elements in general, our devices were optimized based on the unitary Jones matrices to maximize the efficiency of each meta-atom. This discrepancy inevitably introduces crosstalk between the output channels of polarization–OAM-multiplexed holography. Nevertheless, in most cases, the crosstalk is sufficiently suppressed during optimization unless too many target images are encoded into a single device with limited degrees of freedom.

To qualitatively validate the suppression of crosstalk in our devices, we numerically reconstructed intensity images of the optimized design as in Figure S3(a). The target input polarization states are *x*- and *y*-polarizations and target helical mode indices were given as $l$ = 1, 2. The top and bottom rows represent raw intensity and filtered intensity images, respectively. Comparing the columns directly reveals the crosstalk between different OAM and polarization channels (e.g., the comparison between the first and second columns shows the crosstalk between polarization channels with the same mode index, and so on). The filtered results explicitly demonstrate that the crosstalk in our designs was sufficiently suppressed and became negligible after removing unwanted OAM modes with doughnut-like profiles.

*Crosstalk due to experimental imperfections*

Even though optimization reduces crosstalk, experimental imperfections may further degrade device performance. Specifically, unavoidable deviations in structural dimensions and rotation angle of nanoposts during fabrication alter local phase and polarization responses, hindering precise realization of the target Jones matrices. Additionally, misalignment of polarization optics in the optical characterization setup can deteriorate channel separation.

Despite these experimental imperfections, the fabricated samples and measurements confirmed that TAM vectorial holography can still be realized with sufficiently low crosstalk. Figure S3(b) shows the measured intensity images of the VB-multiplexed vectorial holograms, demonstrating experimental robustness. As with the numerical results, comparing the columns reveals crosstalk between polarization and OAM channels. In this case, since VB #1 and VB #2 are from the same HOPS, they indirectly reveal the crosstalk between polarization channels. On the contrary, the vector beam states from different HOPSs can show the crosstalk between OAM channels. Note that the blurred images with non-target OAM modes observed in the raw intensity images (top row) are barely seen in the filtered images (bottom row).

## S5. Post-processing procedure for reconstructing vectorial holographic images

To reconstruct target holographic images selectively for specific OAM modes, an aperture array was applied to the measured images. Ideally, this aperture array is aligned in the $k_x$ and $k_y$ axes. However, due to experimental misalignment of the optical measurement setup, the measured results were generally misaligned (rotated and translated) relative to the aperture array. Therefore, before applying the aperture array, the measured images were rotated and translated to properly match the aperture array, and then the aperture array is applied. Figure S4 illustrates the post-processing procedure for one measured image. The filtered images were subsequently used to calculate the Stokes parameters, as described in a later section.

While we used a preset aperture array to filter out unwanted OAM modes, reconstruction quality can be further improved either through physical calibration or algorithmic compensation of the optical measurement setup. For example, using a reference metasurface with a uniform dot array matched to the aperture array would allow precise registration of aperture positions, thereby enhancing reconstruction accuracy. Alternatively, the aperture array can be algorithmically optimized to account for optical aberrations and distortions inferred from the measured data, enabling compensation for optical system-induced possible imperfections without requiring explicit prior knowledge of these aberrations and distortions.

## S6. Design of input-generating metasurfaces

For the experimental demonstration of the suggested TAM, VB-multiplexed, and bidirectional TAM vectorial holograms, various input states were required. Specifically, two types of input states were necessary: (1) OAM modes with spatially invariant polarization states and (2) vector beams with spatially varying polarization states. To generate these states, two kinds of input state-generating metasurfaces were designed. In the experiment, both types of input state-generating metasurfaces were placed immediately before the hologram samples, as shown in Figure S5.

*Polarization-insensitive OAM mode generation*

We used TAM states with various polarization states that remained spatially invariant within each TAM state. To generate these input states, we employed polarization-insensitive OAM mode-generating metasurfaces, which can generate OAM modes without altering the polarization state of the input beam. The required Jones matrix is given as $J = \begin{bmatrix} \mathrm{e}^{il\varphi} & 0 \\ 0 & \mathrm{e}^{il\varphi} \end{bmatrix}$ where $l$ and $\varphi$ represent helical mode index and azimuthal angle in cylindrical coordinates. This can be easily realized using isotropic nanoposts. With the aid of polarization optics, various TAM states can then be prepared.

*Vector beam generation*

As mentioned in the main text, each vector beam state used in this study is given as $|\psi\rangle = \alpha|p_1, l_1\rangle + \beta|p_2, l_2\rangle$. Since we used two orthogonal polarization states ($\langle p_1|p_2\rangle = 0$) and normalized coefficients ($|\alpha|^2 + |\beta|^2 = 1$) for constructing VB states, the required vector fields can be written as $|\psi\rangle = \begin{bmatrix} \cos\chi\,\mathrm{e}^{i\xi_x} \\ \sin\chi\,\mathrm{e}^{i\xi_y} \end{bmatrix}$ with unity magnitude across the design plane, where $\chi$, $\xi_x$, and $\xi_y$ were uniquely defined by the required polarization and phase distributions. To generate these fields under *x*-polarization illumination, the required Jones matrix is $J = \begin{bmatrix} \cos\chi\,\mathrm{e}^{i\xi_x} & \sin\chi\,\mathrm{e}^{i\xi_y} \\ \sin\chi\,\mathrm{e}^{i\xi_y} & -\cos\chi\,\mathrm{e}^{i(2\xi_y-\xi_x)} \end{bmatrix}$, which is unitary and symmetric. This can be realized using single-layer metasurfaces. Note that a different single-layer metasurface was required for each input vector beam state.

## S7. Optical characterization of the Stokes parameters

The polarization states of light can be represented using the Stokes parameters. The Stokes parameters are defined as

$$\begin{aligned} S_0 &= I_x + I_y = I_{45^\circ} + I_{135^\circ} = I_R + I_L \\ S_1 &= I_x - I_y \\ S_2 &= I_{45^\circ} - I_{135^\circ} \\ S_3 &= I_R - I_L \end{aligned} \qquad \text{(S6)}$$

where $I$ denotes the measured intensity, and the subscripts $x$, $y$, 45°, 135°, $R$, and $L$ represent $x$-polarized, $y$-polarized, 45° linearly polarized, 135° linearly polarized, right-circularly polarized, and left-circularly polarized states. Therefore, these parameters can be quantitatively retrieved by measuring polarized intensity using the optical measurement setup shown in Figure S5. From the Stokes parameters, two spherical coordinates (azimuth and elevation on the Poincaré sphere) can be directly obtained from the following relation given as

$$2\psi = \begin{cases} \tan^{-1}\left(\frac{S_2}{S_1}\right) & (S_1 \geq 0) \\ \tan^{-1}\left(\frac{S_2}{S_1}\right) + \pi & (S_1 < 0) \end{cases}$$

$$2\chi = \tan^{-1}\left(\frac{S_3}{S_0}\right). \qquad \text{(S7)}$$

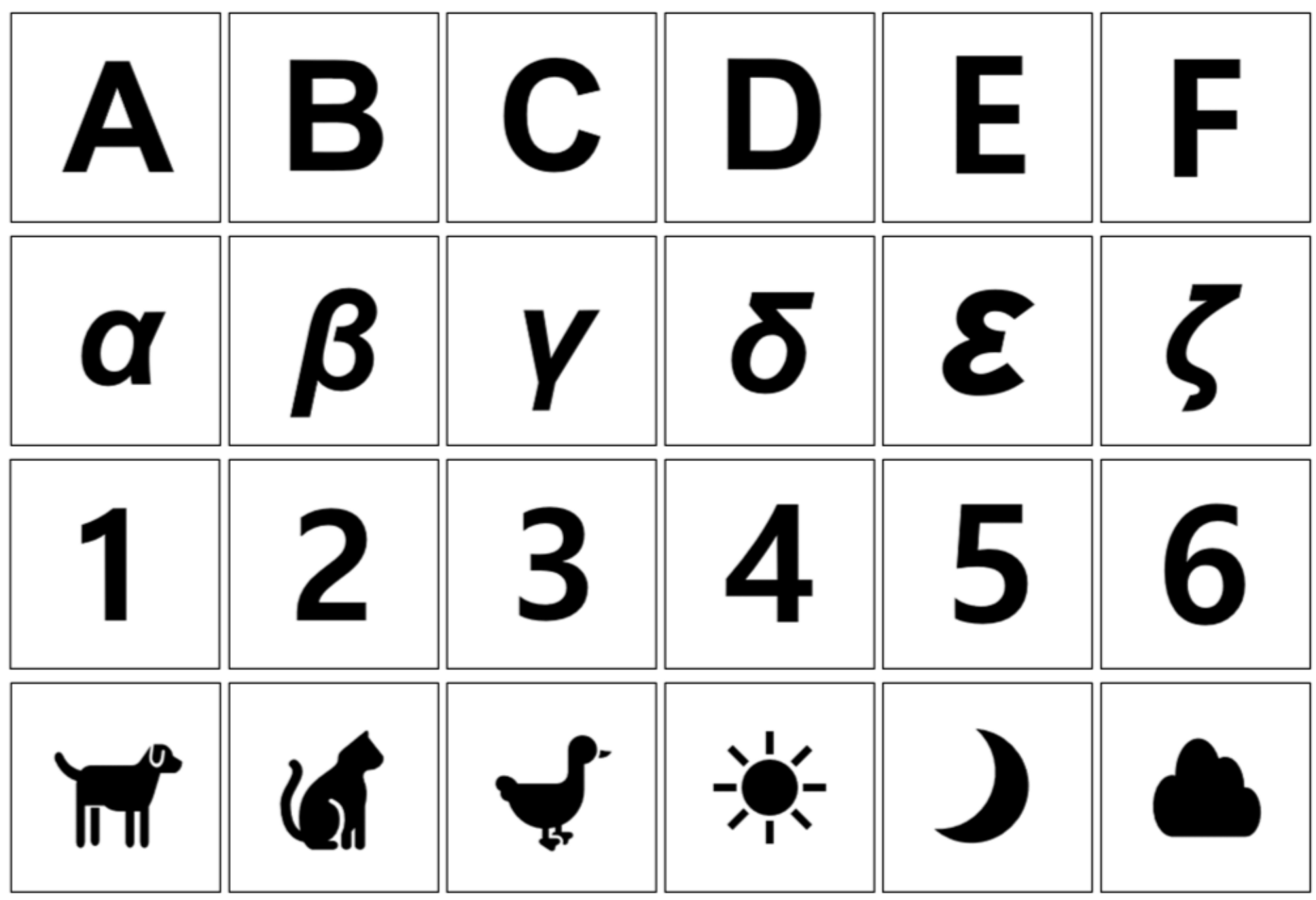

**Figure S1. Ground truth images.**

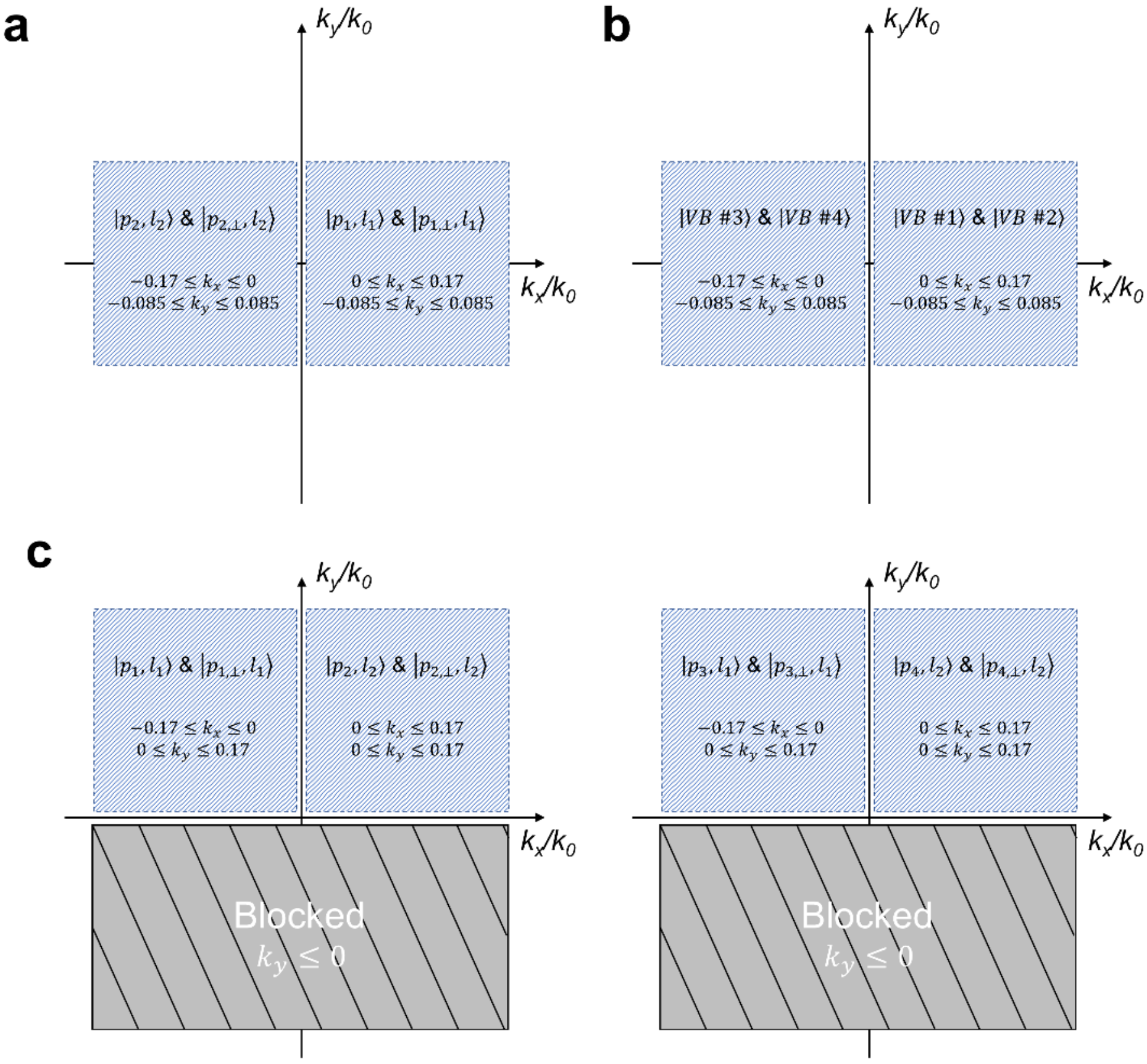

**Figure S2. Spatial separation of output channels in the spatial-frequency domain.** (a) Configuration for TAM vectorial holography. (b) Configuration for VB-multiplexed vectorial holography. (c) Configuration for bidirectional TAM vectorial holography. The right and left panels correspond to the front- and back-side illumination conditions.

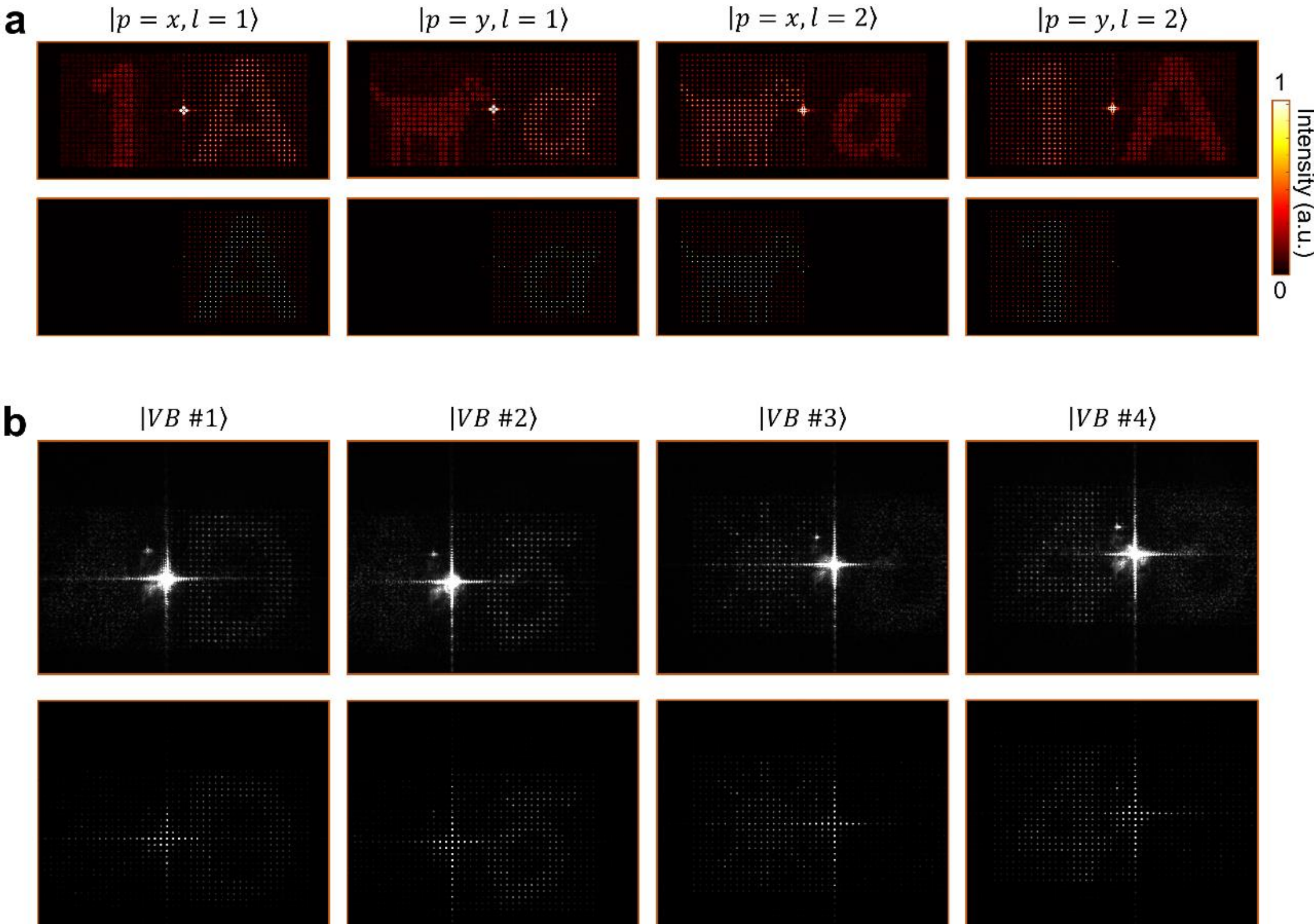

**Figure S3. Crosstalk analysis.** (a) Numerical verification of polarization–OAM channel crosstalk in TAM vectorial holography. Top: raw intensity images; bottom: filtered intensity images for each input TAM state. (b) Experimental verification polarization–OAM channel crosstalk in VB-multiplexed vectorial holography. Top: raw intensity images; bottom: filtered intensity images for each input VB state.

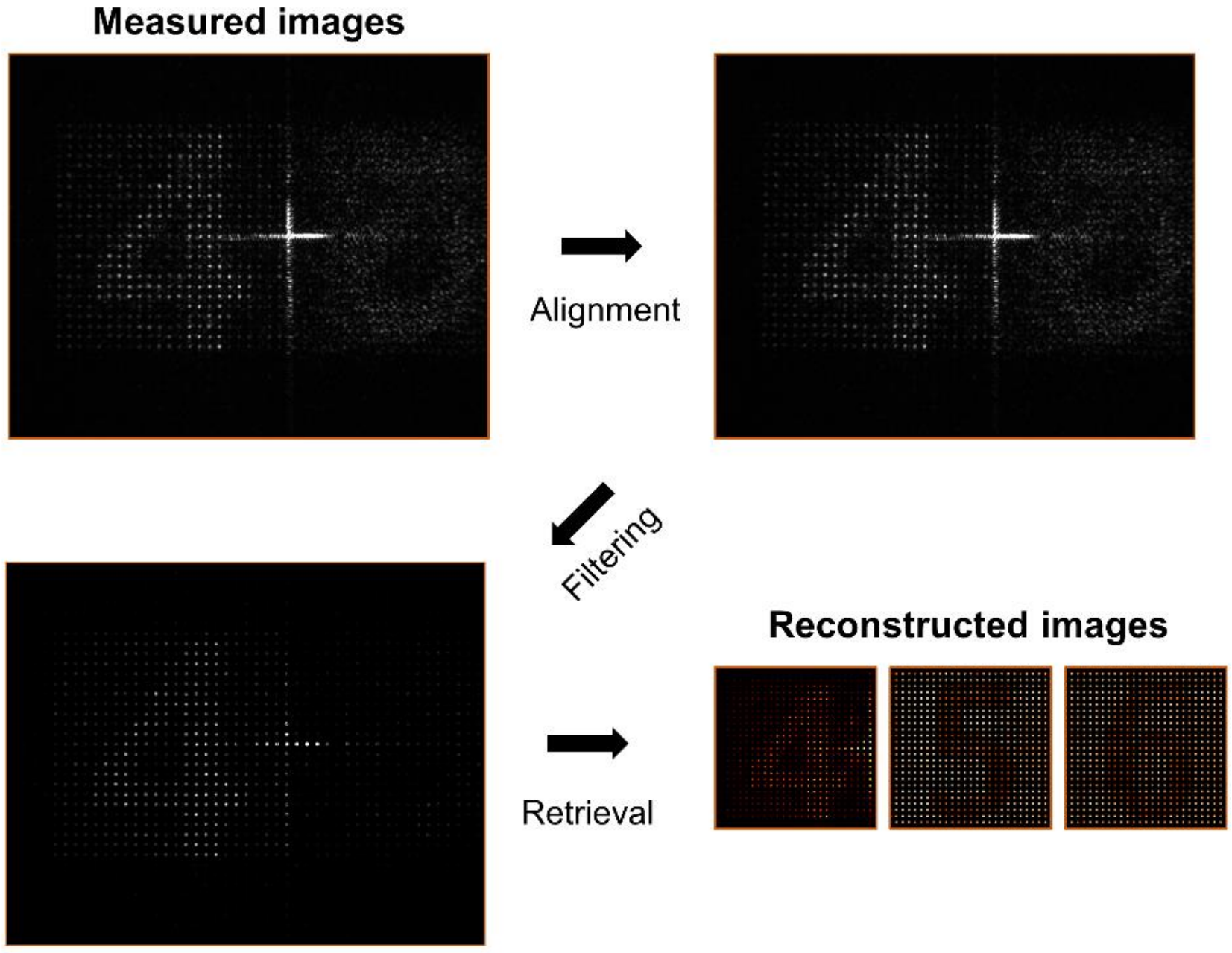

**Figure S4. Post-processing procedure.** Measured results were aligned (rotated and translated), filtered by the aperture array, and then used to retrieve the Stokes parameters.

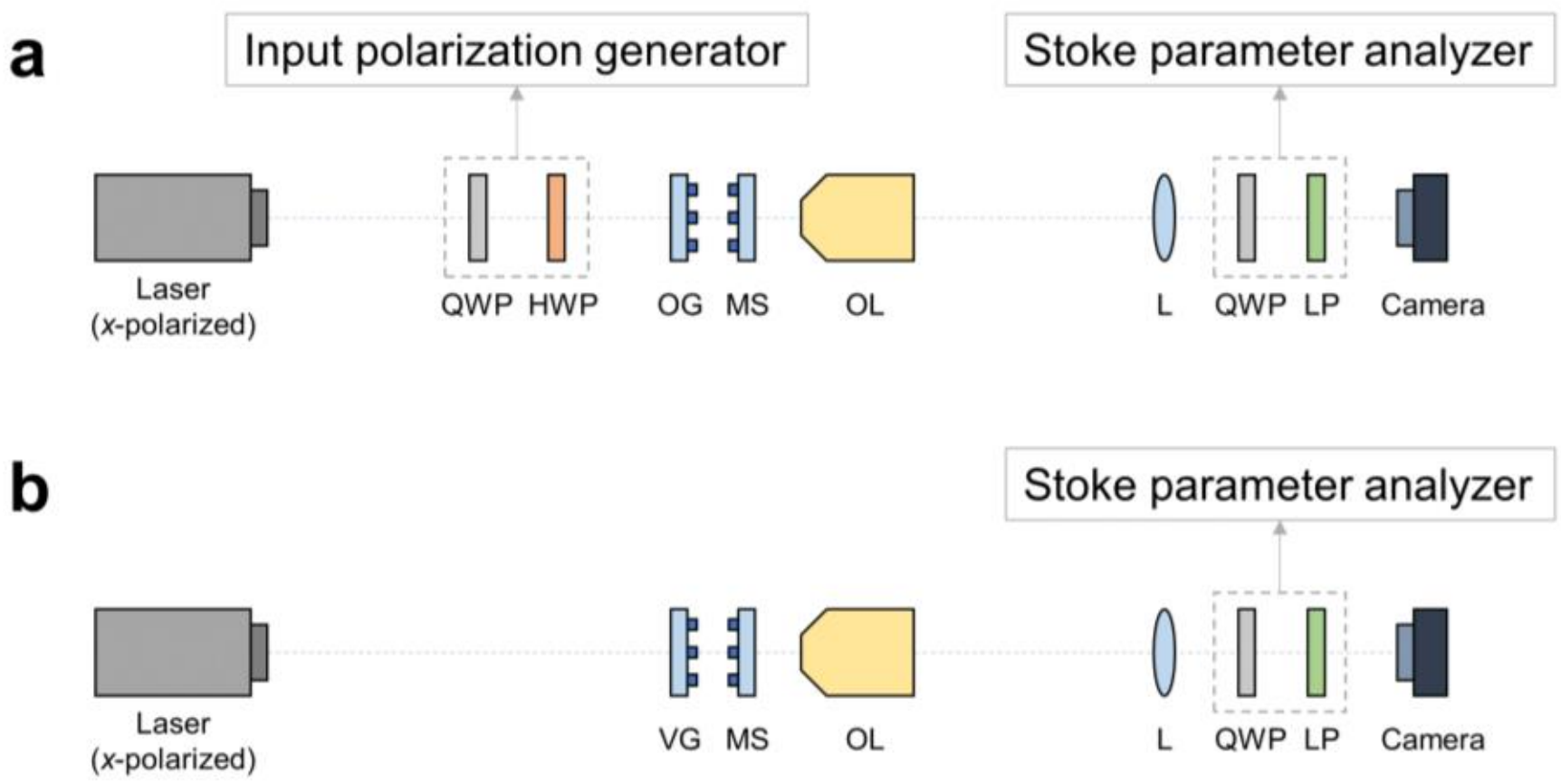

**Figure S5. Optical characterization setup.** Experimental setup for measuring far-field holographic images with (a) TAM states and (b) VBs as input. The holographic images are formed at the back focal plane of the objective lens and relayed to the camera using an additional lens is placed to transfer the images to the camera. QWP: quarter-wave plate; HWP: half-wave plate; MS: metasurfaces; OL: objective lens; L: lens; LP: linear polarizer; OG: OAM input-generating metasurfaces; VG: VB input-generating metasurface.

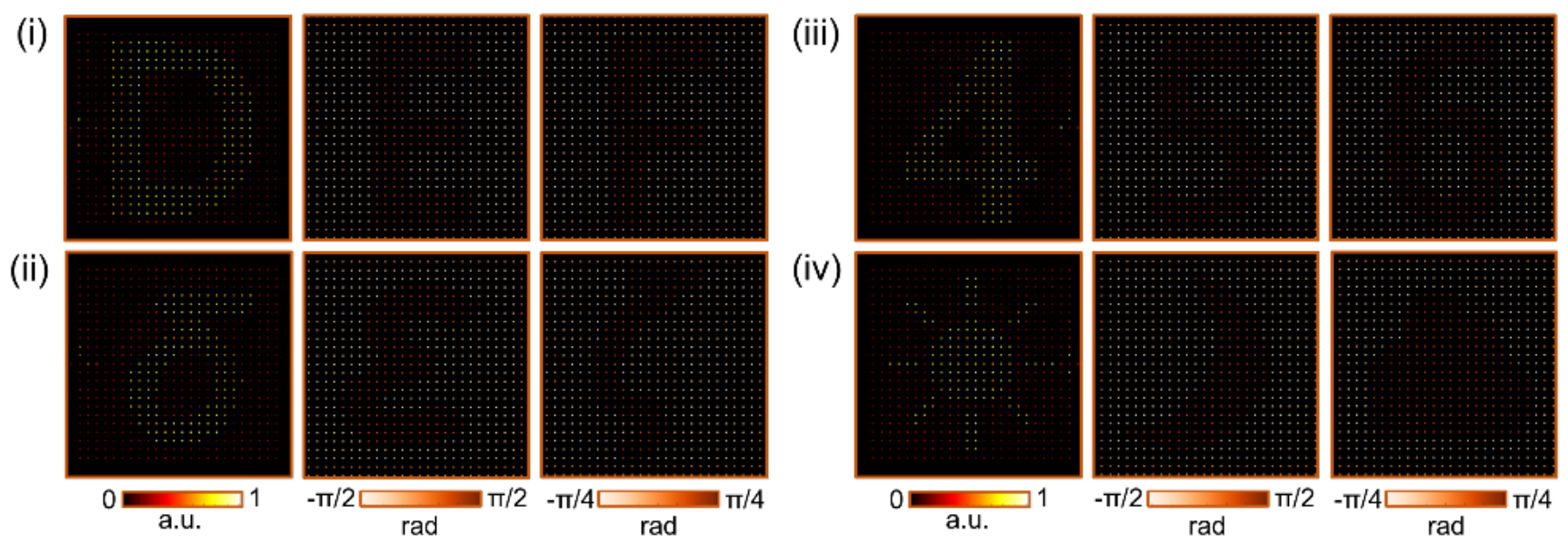

**Figure S6. Numerical demonstration of VB-multiplexed vectorial holography.**

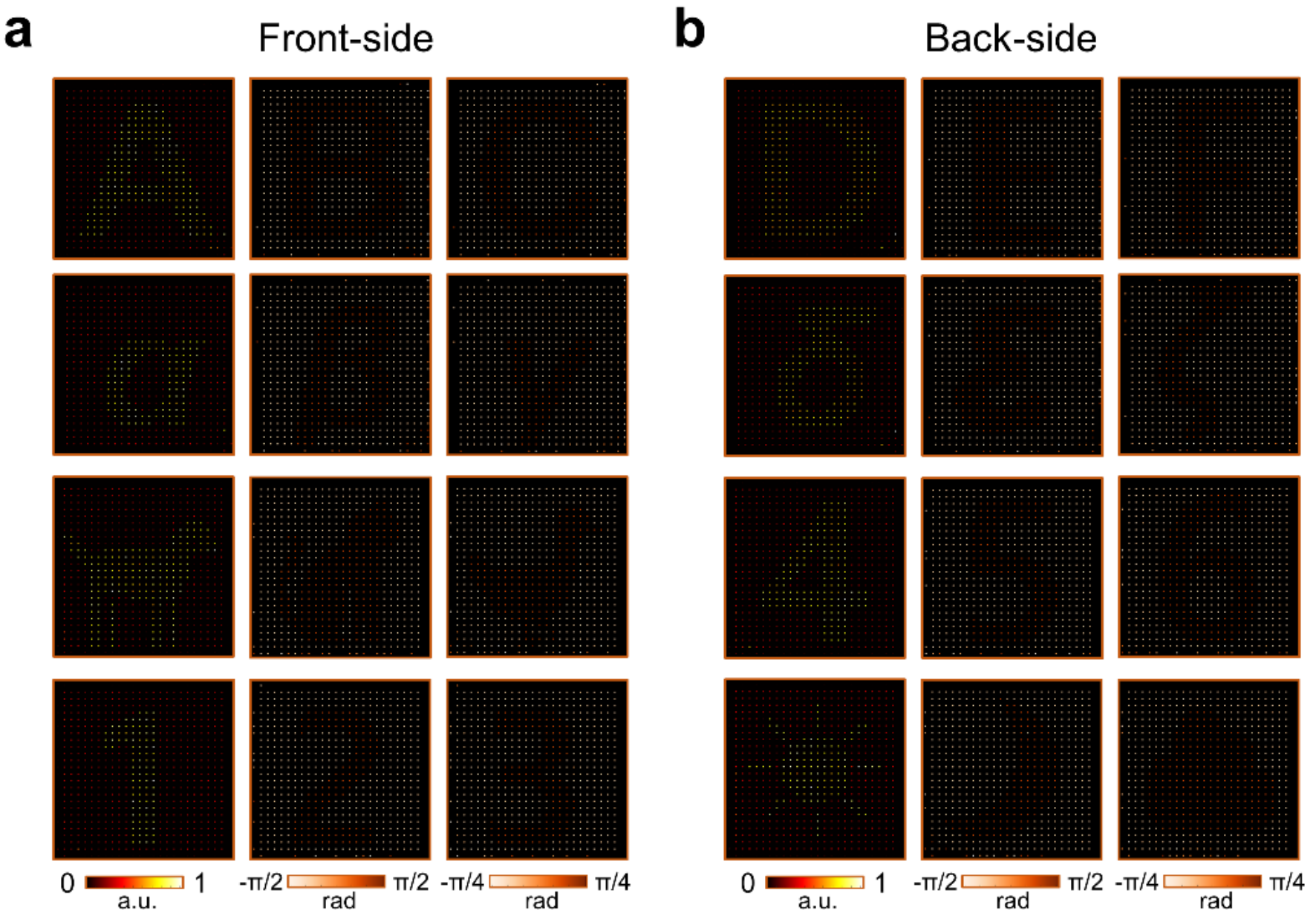

**Figure S7. Numerical demonstration of bidirectional TAM vectorial holography.** (a) Reconstructed images for the front-side illumination. (b) Reconstructed images for the back-side illumination.

## References

1. Menzel, C., C. Rockstuhl, and F. Lederer, *Advanced Jones calculus for the classification of periodic metamaterials.* Physical Review A—Atomic, Molecular, and Optical Physics, 2010. **82**(5): p. 053811.
2. Arbabi, A., et al., *Dielectric metasurfaces for complete control of phase and polarization with subwavelength spatial resolution and high transmission.* Nature nanotechnology, 2015. **10**(11): p. 937-943.
3. Chang, T., et al., *Universal metasurfaces for complete linear control of coherent light transmission.* Advanced Materials, 2022. **34**(44): p. 2204085.
4. Fang, X., H. Ren, and M. Gu, *Orbital angular momentum holography for high-security encryption.* Nature Photonics, 2020. **14**(2): p. 102-108.